\pdfoutput=1

\documentclass[10pt,twocolumn]{article}

\PassOptionsToPackage{colorlinks=true,linkcolor=blue!55!black,%
            citecolor=blue!45!black,urlcolor=blue!55!black}{hyperref}

\usepackage[a4paper,top=2.2cm,bottom=2.4cm,left=1.9cm,right=1.9cm,%
            columnsep=0.6cm]{geometry}
\usepackage[T1]{fontenc}
\usepackage[utf8]{inputenc}
\usepackage{lmodern}
\usepackage{microtype}
\usepackage{graphicx}
\usepackage{amsmath}
\usepackage{amssymb}
\usepackage{booktabs}
\usepackage{threeparttable}
\usepackage[font=small,labelfont=bf]{caption}
\usepackage[round]{natbib}
\usepackage{xcolor}
\usepackage{orcidlink}
\usepackage{authblk}
\usepackage{hyperref}

\graphicspath{{./}}

\newcommand{\doi}[1]{\href{https://doi.org/#1}{doi:#1}}

\hypersetup{
  pdftitle={Circumstellar Disc and X-ray Variability in the Be/X-ray Binary SXP 5.05 During its 2024 Outburst},
  pdfauthor={Chintan Patel, Sayantan Bhattacharya, Karan Akbari, Rajapandi Nadar, Sudip Bhattacharyya, Manojendu Choudhury},
  pdfkeywords={X-ray astronomy, accretion, high-mass X-ray binaries, X-ray pulsars, neutron stars, SXP 5.05}
}

\title{\bfseries Circumstellar Disc and X-ray Variability in the Be/X-ray Binary
SXP~5.05 During its 2024 Outburst}

\author[1]{Chintan Patel\,\orcidlink{0009-0006-8622-5471}}
\author[2]{Sayantan Bhattacharya\,\orcidlink{0000-0001-8572-8241}}
\author[1]{Karan Akbari\,\orcidlink{0009-0005-0550-4018}}
\author[1]{Rajapandi Nadar\,\orcidlink{0009-0006-9067-1132}}
\author[2]{Sudip Bhattacharyya\,\orcidlink{0000-0002-6351-5808}}
\author[1]{Manojendu Choudhury}

\affil[1]{St.~Xavier's College, 5 Mahapalika Marg, Fort, Mumbai 400001, India}
\affil[2]{Department of Astronomy and Astrophysics, Tata Institute of Fundamental
Research, 1 Homi Bhabha Road, Colaba, Mumbai 400005, India}

\date{}

\begin{document}

\twocolumn[
\begin{@twocolumnfalse}
\maketitle
\vspace{-1.5em}
\begin{abstract}
\noindent
Be/X-ray binaries provide a unique opportunity to study the interaction between neutron stars and circumstellar discs. SXP~5.05 is a particularly rare system, exhibiting eclipse-like X-ray variability attributed to obscuration by the Be star disc rather than a simple stellar eclipse. Motivated by its unusual geometry and the well-studied 2013 outburst, we present a multiwavelength analysis of its 2024 outburst using \textit{NICER} X-ray observations and long-term optical monitoring from OGLE. The X-ray light curve shows a declining outburst with lower peak intensity and shorter duration compared to 2013, indicating a reduced accretion episode. The spectral evolution, characterized through hardness ratios, reveals a transition from a soft, high-intensity state to a harder, low-intensity state. Coherent pulsations near 5.05~s are detected throughout the observations, with properties consistent with previous measurements. The optical light curves show a reduced variability amplitude relative to 2013, possibly from a less extended or less dense circumstellar disc. Orbital-phase-folded optical profiles reveal a persistent, phase-locked dip structure, indicating a stable non-axisymmetric disc component that evolves across outburst phases. Together, these results support a picture in which the observed variability is driven by changes in disc structure and viewing geometry. SXP~5.05 thus remains a key system for probing the time-dependent properties of Be star discs through combined X-ray and optical observations.
\end{abstract}

\vspace{0.4em}
\vspace{1.2em}
\end{@twocolumnfalse}
]

\section{Introduction}\label{sec:intro}

Be/X-ray binaries (BeXRBs) form the largest subclass of high-mass X-ray binaries and consist of a neutron star in orbit around a rapidly rotating Be-type star surrounded by a circumstellar decretion disc \citep{Reig2011}. The X-ray emission in these systems is powered by accretion of material from the Be star disc onto the neutron star, typically occurring during periastron passages or during phases of enhanced disc growth. These systems exhibit a wide range of variability, including periodic outbursts and long-term changes linked to the formation, evolution, and dissipation of the circumstellar disc.

The Small Magellanic Cloud (SMC) hosts a rich population of BeXRBs, providing a unique laboratory for studying accretion processes and disc--neutron star interactions under relatively uniform distance and low foreground absorption \citep{Coe2005}. Among these systems, SXP~5.05 is particularly notable due to its unusual eclipsing behavior, first reported during the 2013 outburst \citep{previous_spin}. Unlike typical BeXRBs, where variability is primarily driven by changes in accretion rate, SXP~5.05 exhibits periodic X-ray dips that were interpreted as obscuration by the circumstellar disc of the Be star rather than a simple stellar eclipse.

The 2013 outburst of SXP~5.05 provided the first detailed insight into this behavior. The system displayed complex, energy-dependent X-ray variability, including deep absorption features and eclipse-like structures occurring at specific orbital phases \citep{previous_spin}. These features were interpreted as the neutron star moving through or behind a warped and non-axisymmetric circumstellar disc. Subsequent modelling of the system suggested that the disc structure plays a dominant role in shaping the observed X-ray properties, with variations in density, scale height, and geometry leading to changes in both flux and absorption \citep{Brown2019}.

SXP~5.05 remains one of the few BeXRBs where disc occultation signatures can be directly studied through multiwavelength observations. Such systems are of particular interest because they allow the neutron star to act as a probe of the disc structure. As the neutron star orbits the Be star, it samples different regions of the disc, providing information on its density distribution, geometry, and temporal evolution. In this context, multiwavelength observations are essential, as optical emission traces the size and density of the disc, while X-ray observations probe the accretion flow and line-of-sight absorption.

In this work, we present a detailed study of the 2024 outburst of SXP~5.05 using observations from the \textit{Neutron Star Interior Composition Explorer} (\textit{NICER}), complemented by long-term optical monitoring from the Optical Gravitational Lensing Experiment (OGLE). We investigate the spectral and timing properties of the source, examine the evolution of its optical light curves, and compare the 2024 outburst with the well-studied 2013 event.

Our goal is to understand how the structure of the circumstellar disc influences the observed X-ray and optical variability, and to explore how the properties of the system evolve between different outburst epochs. In particular, we focus on the connection between optical circumstellar disc signatures and X-ray spectral evolution, building on the framework established by previous studies of this system.

The article is structured as follows: Section~\ref{sec:intro} starts with a general motivation and provides the particular motivation for this work. Section~\ref{sec:DA} describes the data analysis and details of the observations used. Results are shown in Section~\ref{sec:res} and discussed in Section~\ref{sec:discussion}. Finally, we summarize in Section~\ref{sec:conc}.

\section{Observations and Data Analysis}\label{sec:DA}
\subsection{X-ray: NICER}\label{sec:xray-nicer}

\begin{figure*}[!t]
 \centering
 \includegraphics[width=0.72\textwidth]{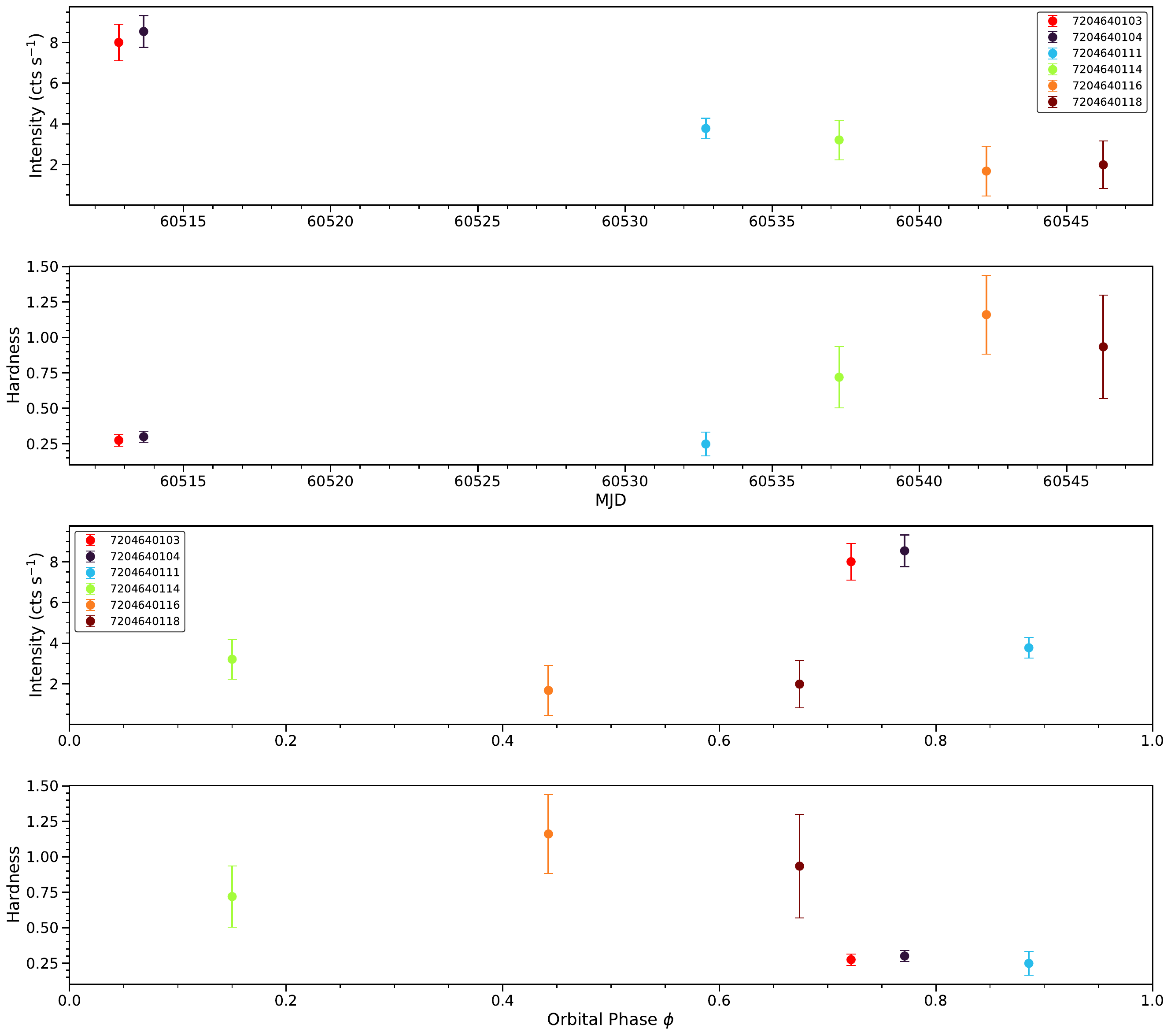}
 \caption{Evolution of the intensity and hardness of SXP~5.05 during the outburst. The \textit{top two panels} show the intensity and hardness, respectively, as a function of time (MJD), while the \textit{bottom two panels} show the same quantities as a function of orbital phase. The source evolves from a bright, soft state toward lower intensity and higher hardness as the outburst progresses. No subsequent softening is observed within the available coverage, suggesting either that the transition was not monitored or that the observed evolution is affected by additional effects related to the geometry and obscuration of SXP~5.05 (see Section~\ref{sec:discussion} for details).}
 \label{fig:Figure1}
\end{figure*}

NICER \citep{NICER_telescope}, launched in 2017, is an X-ray timing and spectroscopy payload aboard the International Space Station dedicated to the study of neutron stars in the soft X-ray band. The primary science instrument, the X-ray Timing Instrument (XTI; \citealt{NICER+XTI}), operates over the 0.2--12~keV energy range and consists of an array of 56 co-aligned X-ray concentrator optics coupled to silicon drift detectors. This design provides high throughput together with excellent spectral sensitivity and sub-microsecond absolute timing precision, enabling detailed pulse-phase-resolved studies of accreting neutron stars. Photon arrival times are referenced using the onboard Global Positioning System receiver, achieving an absolute timing accuracy better than 300~ns. SXP~5.05 was monitored with NICER from 2024 July 5 to 2024 August 24. For our analysis, we considered 18 observations (ObsIDs 7204640101--7204640118). After standard pipeline reduction and GTI filtering, we found that ObsIDs 7204640101 and 7204640105--7204640110 yielded event files with empty GTI intervals, i.e., no valid photon events were present despite the listed exposure times. These datasets were discarded from further analysis. A complete log of all observations is provided in Table~\ref{table1}, where the excluded ObsIDs are also marked. To process the data and generate event files for each observation, we used the standard \texttt{nicerl2} pipeline included in the \texttt{HEASOFT v6.36} software package, along with the standard NICER analysis tools. For generating light curves and spectra for each observation, we used the \texttt{nicerl3-lc} and \texttt{nicerl3-spect} pipelines, respectively, with the background model \texttt{SCORPEON v23} \citep{SCORPEON}. To ensure that the light curves and hardness--intensity diagrams (HIDs) were not contaminated by non-X-ray flares, we reran the \texttt{nicerl2} pipeline using \texttt{cor\_range="2.0-*"}, since precipitation flares are almost always limited to intervals with $\mathrm{COR\_SAX} < 1.5$. After filtering out these non-source flaring events, we re-extracted the light curves and spectral products using the above tools for further analysis.

NICER is a non-imaging instrument: every source inside the collimator aperture contributes to the recorded counts. The XTI accepts photons from a circular sky region of radius $\sim$3.2~arcmin (31.6~arcmin$^2$; \citealp{remillard2022}), and its off-axis response is flat within $\sim$2~arcmin of the pointing and drops sharply beyond $\sim$3~arcmin, with tabulated fractions of 0.87 at 2.1~arcmin and 0.07 at 4.2~arcmin \citep[their fig.~7 and table~7]{bogdanov2019}. To quantify the field contribution we followed \citet{bogdanov2019}: we took all sources within 10~arcmin of SXP~5.05 from the 4XMM-DR13 catalogue \citep[][VizieR IX/69]{webb2020}, weighted each 0.5--12~keV catalogue flux by the off-axis response at its separation, and converted the weighted fluxes to NICER count rates using the measured flux-to-rate ratio of SXP~5.05 itself ($4.0\times10^{-11}$~erg~cm$^{-2}$~s$^{-1}$ at $\sim$9~counts~s$^{-1}$; Table~\ref{tab:specfit}).

The known Be/X-ray binaries in this field (SXP~7.92, SXP~140, SXP~565, SXP~293, SXP~645, SXP~59.0) lie 5.5--9.5~arcmin from SXP~5.05, outside the response. Of the sources labelled in fig.~3 of \citet{previous_spin}, only the $z=0.15$ quasar (their source C, 1.4~arcmin away) and the HMXB candidate XMMU~J005724.0$-$722357 (their source D, 2.6~arcmin away) fall inside it. Six further catalogued sources within 2.2~arcmin are artifacts of the 2013 outburst itself: each is a single detection in one of the three 2013 XMM-Newton ToOs (ObsIDs 0700580101, 0700580401, 0700580601), each carries the 4XMM manual screening flag, and none appears in the deeper SMC survey of \citet{sturm2013} or in eRASS1 \citep{merloni2024}. Excluding them, the field sums to $\simeq$0.16--0.19~counts~s$^{-1}$, of which 0.12~counts~s$^{-1}$ comes from the quasar alone (the quoted range reflects one ambiguous detection 4.8~arcsec from the target, consistent with quiescent SXP~5.05 itself). The conversion assumes the spectral shape of SXP~5.05 for every field source, and folding softer AGN-like spectra through the NICER effective area moves individual rates by up to a factor of $\sim$3. This contribution is 2--3 per cent of the source rate for the two bright fitted spectra (ObsIDs 7204640103 and 7204640104) and does not alter the HID morphology, but it reaches $\sim$13 and $\sim$22 per cent for the two faintest fitted spectra (ObsIDs 7204640111 and 7204640114), comparable to their statistical uncertainties. The pulse detections are unaffected because field sources add only unpulsed counts; the pulsed fractions in Table~\ref{tab:gti_timing}, all measured from the two bright observations, are diluted by 2--3 per cent at most.

XMMU~J005724.0$-$722357 needs a separate note. In the living Swift-XRT point-source catalogue \citep{evans2023} it reached its brightest recorded XRT count rate, $\sim$$4\times10^{-2}$~counts~s$^{-1}$ ($L_X$ of order $10^{37}$~erg~s$^{-1}$ at the SMC distance), on MJD~60556, ten days after our last NICER observation, and the public XRT coverage has no epoch inside MJD~60496--60546, so its state during our campaign is unconstrained. Its catalogued mean flux implies only 0.01~counts~s$^{-1}$ in NICER, but at the MJD~60556 level it would have added 0.2--0.3~counts~s$^{-1}$ to our last observations; the faintest points in Fig.~\ref{fig:Figure1} carry this caveat.

\begin{table}[!t]
\caption{NICER observation log of SXP~5.05 used in this work. Marker symbols indicate observations excluded from further spectral analysis for specific reasons.}
\label{table1}
\centering
\small
\begin{tabular}{ccc}
\toprule
ObsID & Exposure (s) & MJD \\
\midrule
7204640101$^{\dagger}$ & 534  & 60496.80601 \\
7204640102$^{\dagger}$ & 1309 & 60496.99929 \\
7204640103             & 572  & 60512.72662 \\
7204640104             & 1722 & 60513.24421 \\
7204640105$^{\dagger}$ & 657  & 60522.73495 \\
7204640106$^{\dagger}$ & 2852 & 60525.64560 \\
7204640107$^{\dagger}$ & 2162 & 60527.58194 \\
7204640108$^{\dagger}$ & 299  & 60528.03380 \\
7204640109$^{\dagger}$ & 673  & 60529.90301 \\
7204640110$^{\dagger}$ & 1753 & 60530.03264 \\
7204640111             & 551  & 60532.73150 \\
7204640112$^{\ast}$    & 757  & 60534.78935 \\
7204640113$^{\triangle}$ & 313 & 60535.56319 \\
7204640114             & 2106 & 60537.17616 \\
7204640115$^{\P}$      & 2272 & 60540.46412 \\
7204640116$^{\P}$      & 4254 & 60542.20556 \\
7204640117             & 784  & 60544.20231 \\
7204640118$^{\P}$      & 1137 & 60546.20579 \\
\bottomrule
\end{tabular}

\vspace{1mm}
\begin{minipage}{0.92\columnwidth}
\footnotesize
$^{\dagger}$ No events in the GTI after filtering.\\
$^{\ast}$ Too much contamination from non-X-ray origins.\\
$^{\triangle}$ Very small event time after filtering ($\sim 20$~s).\\
$^{\P}$ Spectra are background dominated.
\end{minipage}
\end{table}

\begin{figure}[!t]
 \centering
 \includegraphics[width=\columnwidth]{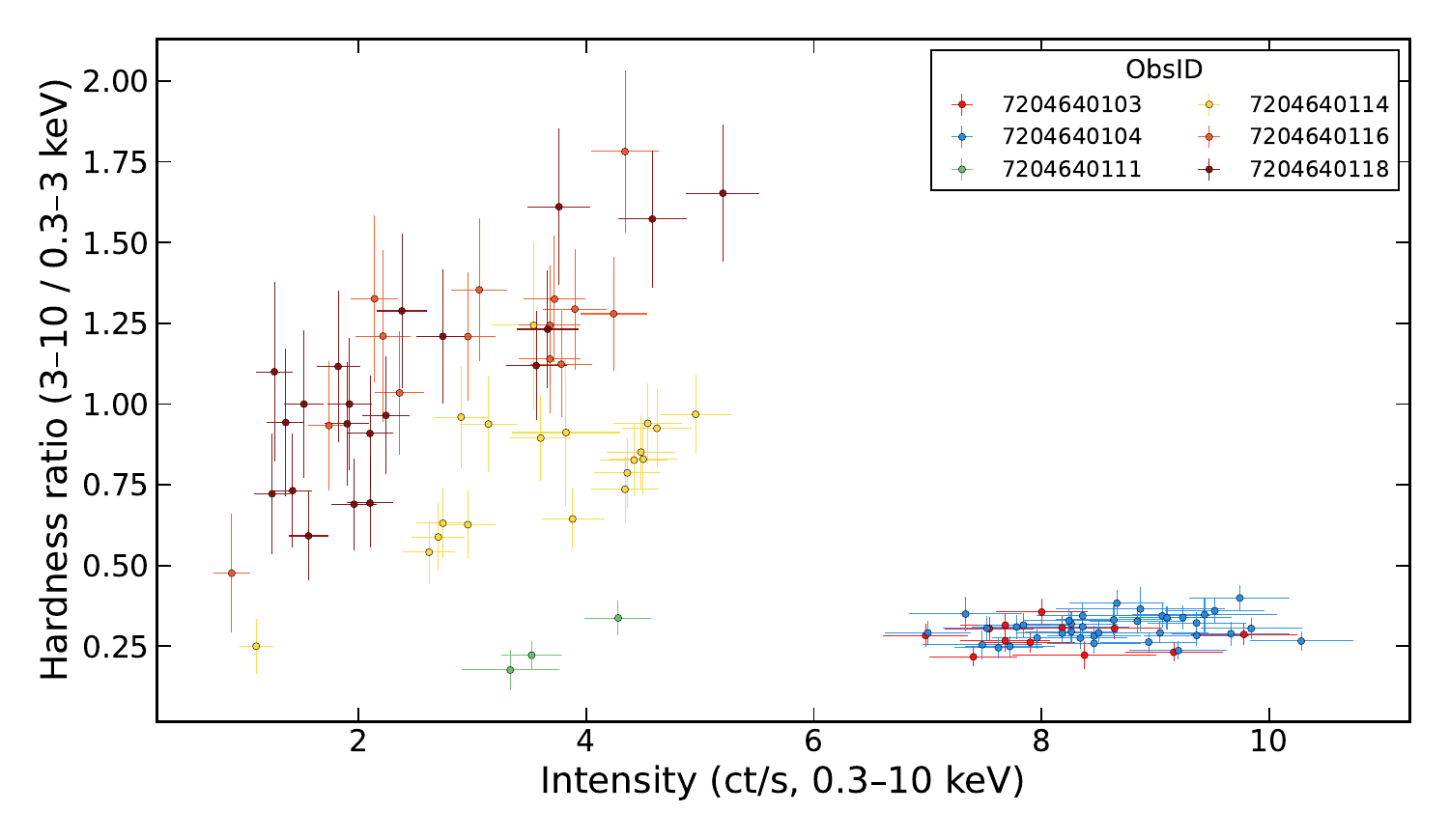}
 \caption{The hardness--intensity diagram (HID) of SXP~5.05 from the NICER observations, with hardness defined as the ratio of count rates in the 3--10~keV and 0.3--3~keV bands. A clear separation is seen in the hardness--intensity plane.}
 \label{Figure2}
\end{figure}

\subsection{Optical Data Analysis}
\label{sec:optical-data-analysis}
The long-term optical behaviour of SXP~5.05 was investigated using $I$-band photometric observations from the Optical Gravitational Lensing Experiment (OGLE-II, OGLE-III and OGLE-IV; \citealt{udalski1997optical,udalski2004optical}) obtained through the X-Ray Variables OGLE Monitoring (XROM) programme, which provides continuous $I$-band monitoring of the optical counterparts of X-ray sources in the OGLE fields, including a large population of SMC X-ray pulsars \citep{udalski2008xromrcomnewogleiii}. The XROM photometry is coarsely calibrated, with typical zero-point uncertainties of $\pm0.1$--$0.2$~mag \citep{udalski2008xromrcomnewogleiii}. For SXP~5.05, the complete XROM light curve spans MJD 50627--60848 (1997 June -- 2025 June), a temporal baseline of $\approx$28~years, consisting of observation times, $I$-band magnitudes, and their photometric uncertainties.

Since the optical light curve exhibits pronounced long-term variability on timescales much longer than the orbital period, it was detrended using the \texttt{flatten} function of the \texttt{wotan} package \citep{hippke2019wotan}, applying a running Tukey biweight filter with a window length of $L=100$~d ($\gg P_{\rm orb}=17.13$~d) so as to remove the slow trend while preserving orbital-scale variability. Within each sliding window $W_i=[t_i-L/2,\,t_i+L/2]$, the local trend was estimated as the biweight-weighted mean $T(t_i)=\sum_j w_j m_j/\sum_j w_j$ \citep{hippke2019wotan}, and the detrended magnitude obtained as $m_{\rm flat}=m-T(t_i)$. Only measurements with finite magnitudes and positive photometric uncertainties were retained, and the filtering was applied independently to continuous observing segments separated by seasonal gaps.

The detrended light curve was phase-folded on the orbital ephemeris, adopting $P_{\rm orb}=17.13$~d and $T_{0}=56680.45$~MJD from \citet{previous_spin}, with orbital phase computed as
\begin{equation}
\phi=\left(\frac{t-T_{0}}{P_{\rm orb}}\right) \bmod 1, \qquad \phi\in[0,1).
\label{eq:phase}
\end{equation}
To construct the folded profile, points were grouped into $N_{\rm bin}=64$ equal-width phase bins, and a weighted mean magnitude was computed within each. For a bin containing $n$ points with flattened magnitudes $m_i$ and errors $\sigma_i$, each point was weighted by $w_i=\sigma_i^{-2}$, giving the weighted mean and its formally propagated error \citep{bevington2003data},
\begin{equation}
\bar{m} = \frac{\sum_i w_i m_i}{\sum_i w_i}, \qquad
\sigma_{\rm formal} = \left(\sum_i w_i\right)^{-1/2}.
\end{equation}
Since unmodelled systematics (residual trends, underestimated XROM zero-points) can inflate the true scatter beyond this formal expectation, the reduced chi-square $\chi^{2}_{\nu} = (n-1)^{-1}\sum_i w_i(m_i-\bar{m})^2$ was used to rescale the error whenever $\chi^2_\nu>1$, following standard practice \citep{bevington2003data,pontzuckerqueloz2006}: $\sigma_{\rm corr} = \sigma_{\rm formal}\sqrt{\max(1,\chi^{2}_{\nu})}$. This was compared against the empirical standard error of the mean, $\sigma_{\rm SEM}=s/\sqrt{n}$ with $s^2=(n-1)^{-1}\sum_i(m_i-\bar{m})^2$, and the more conservative of the two,
$\sigma_{\rm bin}=\max(\sigma_{\rm corr},\sigma_{\rm SEM})$,
was adopted as the final bin uncertainty (for $n=1$, $\sigma_i$ was used directly). This ensures the plotted error bars reflect the true point-to-point dispersion rather than only the nominal, coarsely calibrated XROM errors \citep{udalski2008xromrcomnewogleiii}.

\begin{figure*}[!t]
    \centering
    \includegraphics[width=0.82\textwidth]{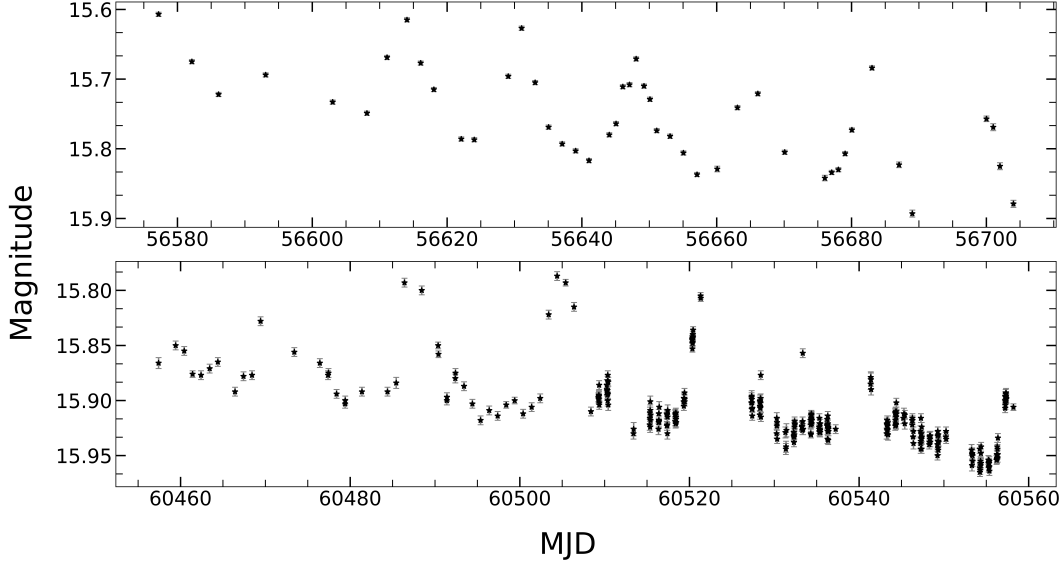}
    \caption{OGLE \textit{I}-band optical light curves of SXP~5.05 during two outburst epochs separated by $\sim$3880 days ($\approx$10.6~yr). \textbf{Top}: MJD 56570--56710 (2013--2014), peak brightness $I \approx 15.60$~mag, amplitude $\Delta I \approx 0.30$~mag. \textbf{Bottom}: MJD 60455--60565 (2024), amplitude $\Delta I \approx 0.15$~mag. This provides a direct comparison of the two outbursts and shows that $\Delta I$ is lower for the 2024 outburst.}
    \label{fig:optical_outburst_lc}
\end{figure*}

This phase-folding and binning procedure was applied independently to the two optical outburst intervals identified in the detrended light curve (Fig.~\ref{fig:optical_outburst_lc}), allowing a direct comparison of the orbital modulation during each activity state. The resulting phase-folded light curves are presented in Fig.~\ref{fig:optical_phase_panels}.

\section{Results}\label{sec:res}
\subsection{Hardness--Intensity Diagram}
\label{subsec:HID}
We extract the 50~s binned light curve for all the observations in the 0.3--10~keV energy range for each NICER observation segment using \texttt{XSELECT}. We depict the evolution of the source in Figure~\ref{fig:Figure1}: the count rate decreases from $\sim$10~counts~s$^{-1}$ to $\sim$1~counts~s$^{-1}$ as the outburst progresses. To investigate whether the change in count rate is associated with a change in the hardness, we plot the hardness versus time (Figure~\ref{fig:Figure1}) and the hardness--intensity diagram (HID; Figure~\ref{Figure2}). The soft band and the hard band for the HID are 0.3--3.0~keV and 3.0--10.0~keV, respectively. The source starts from a very soft state (first two observations), then the intensity drops from $\sim$8--10~counts~s$^{-1}$ to $\sim$3--4~counts~s$^{-1}$ while the hardness remains the same, then progressively the hardness increases and the count rate drops even further. The source evolves from a softer spectrum at high intensity to a harder spectrum at lower intensity.

\subsection{Spectral Analysis}\label{spectra_res}
The \textit{NICER} observations were mostly of short duration; hence, enough statistics to perform a spectral fit were only available for four observations, as shown in Table~\ref{tab:specfit}. Here, we describe the spectral fitting for the data of ObsID 7204640104, as it has a high count rate and one of the longest exposure times among all the observations, but we have applied similar modelling to all the NICER spectra. For spectral fitting, we use the software \texttt{XSPEC v12.15.1} and use $\chi^2$ statistics. We report the 1$\sigma$ uncertainty in the parameters for all the observations. To account for the absorption by the interstellar and intrinsic neutral hydrogen medium, we use the \texttt{XSPEC} model \texttt{tbabs}. We adopt the abundances of \citet{wilms_tbabs} and the cross sections of \citet{verner_cross_section} to model the spectrum. We limit the spectrum for each observation to an energy range where the source counts are above the background counts.

We model the spectra with \texttt{tbabs*(bbodyrad)} and add a \texttt{powerlaw} component to account for non-thermal emission, resulting in $\chi^2$/degrees of freedom (dof) $= 129.60/106$, corresponding to a null hypothesis probability of $5.95\times10^{-2}$. The equivalent hydrogen column density is found to be $N_{\rm H} = (3.05 \pm 0.24)\times10^{21}\,\mathrm{cm^{-2}}$. The thermal component is characterized by a blackbody temperature of $kT = 1.40 \pm 0.31$~keV with a normalization of $0.17 \pm 0.08$. The non-thermal continuum is described by a relatively hard power law with photon index $\Gamma = 1.40 \pm 0.14$. The residuals remain largely confined within $\pm 3\sigma$, indicating that the model reproduces the observed continuum shape without requiring additional spectral components.

\begin{figure}[!t]
 \centering
 \includegraphics[width=\columnwidth]{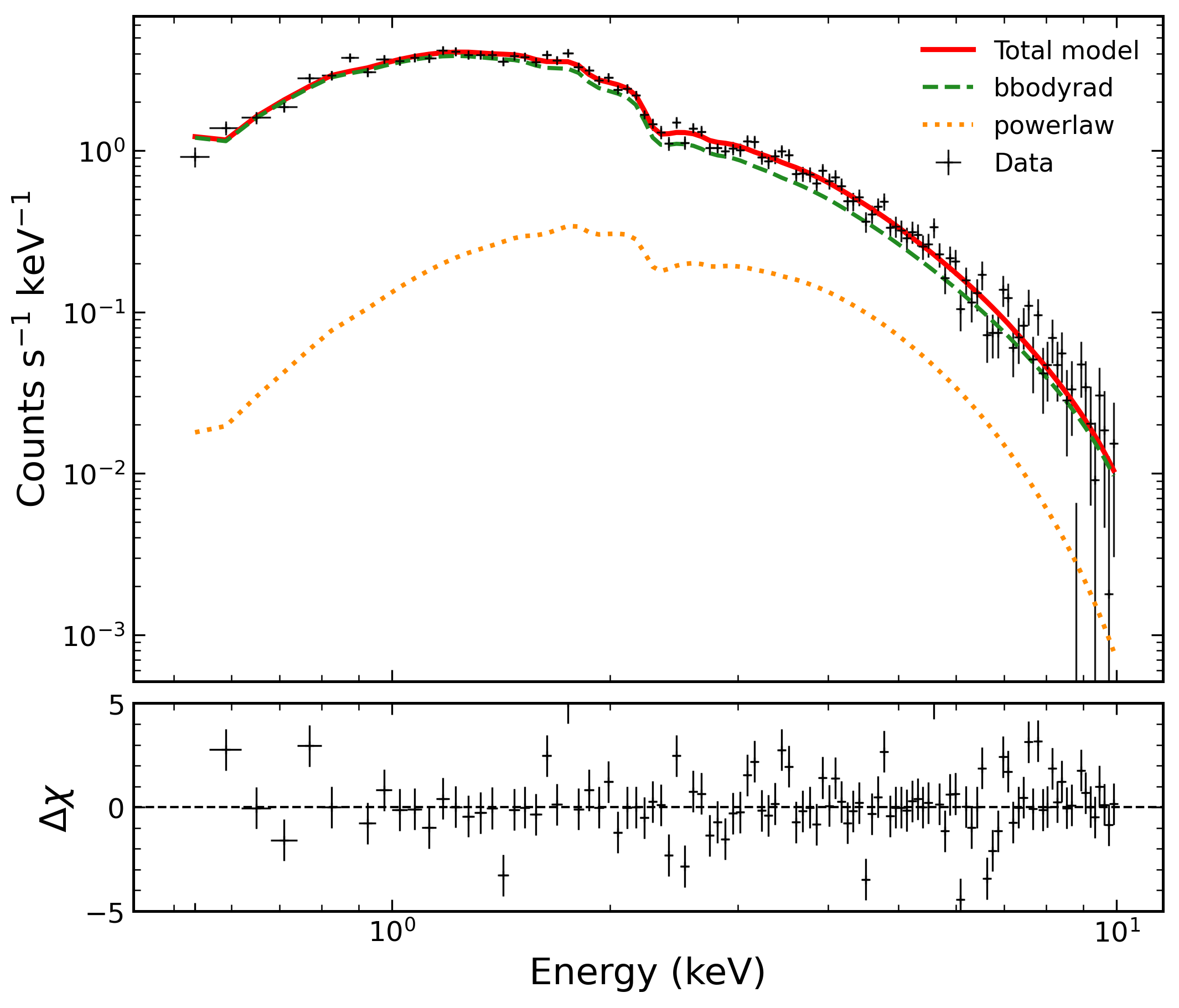}
 \caption{Spectral fitting of the 0.5--10~keV NICER spectrum of SXP~5.05 using the XSPEC model \texttt{TBabs(bbodyrad + powerlaw)} (see Section~\ref{spectra_res}). The top panel shows the unfolded source spectrum (black crosses) together with the best-fitting total model (solid red line). The dashed green curve represents the thermal blackbody component, while the dotted orange curve denotes the non-thermal power-law continuum. The fit yields $\chi^2/\mathrm{dof}=129.6/106$, indicating an acceptable description of the continuum. The lower panel shows the $\Delta\chi$ residuals, which remain largely within $\pm 3\sigma$, demonstrating that the model adequately reproduces the observed spectral shape without requiring additional line or reflection components.}
 \label{Figure4}
\end{figure}

The spectral components may arise from distinct emission regions. The high-temperature blackbody component ($kT \sim 1.4$~keV) most likely originates from a localized hot spot or accretion column base near the neutron-star magnetic poles, where the inflowing material is thermalized at the stellar surface. Such temperatures are typical of polar-cap emission in accreting X-ray pulsars and are significantly hotter than expected from a standard accretion disk.

The hard power-law tail ($\Gamma \sim 1.4$) indicates the presence of Comptonized or non-thermal emission from the accretion column, where soft seed photons are up-scattered by energetic electrons in the magnetized plasma. The fit suggests that the blackbody component dominates the soft X-ray band, while the power law contributes increasingly toward higher energies, consistent with the standard picture of magnetically channeled accretion in Be/X-ray pulsars. The absence of any strong systematic residuals or line-like features suggests that the source shows no statistically significant evidence for additional reflection, iron-line emission, or cyclotron absorption features within the 0.5--10~keV band at the achieved signal-to-noise ratio.

\begin{table*}[!t]
\centering
\begin{threeparttable}
\caption{Best-fit spectral parameters of SXP~5.05 obtained using the XSPEC model
\texttt{TBabs(bbodyrad + powerlaw)}. These are the only observations where significant counts were available for even a simple spectral fit. Errors are quoted at the 90\% confidence level.}
\label{tab:specfit}
\small
\setlength{\tabcolsep}{5.5pt}
\renewcommand{\arraystretch}{1.25}
\begin{tabular}{ccccccccc}
\toprule
ObsID &
TBabs &
\multicolumn{2}{c}{bbodyrad} &
\multicolumn{2}{c}{powerlaw} &
\multicolumn{2}{c}{Flux\tnote{a}} &
$\chi^2/\nu$ \\
\cmidrule(lr){2-2}\cmidrule(lr){3-4}\cmidrule(lr){5-6}\cmidrule(lr){7-8}
 & $N_{\rm H}$ & $kT$ & Norm & $\Gamma$ & Norm\tnote{b} & Abs. & Unabs. & \\
 & ($10^{22}\,\mathrm{cm^{-2}}$) & (keV) & & & ($\times10^{-3}$) &
\multicolumn{2}{c}{($10^{-12}\,\mathrm{erg\,cm^{-2}\,s^{-1}}$)} & \\
\midrule
7204640103 & $0.51\pm0.08$ & $1.24\pm0.07$ & $0.89\pm0.29$ & $2.80\pm0.53$ & $7.0\pm1.6$ & 29.62 & 44.19 & 79.5/79 \\
7204640104 & $0.30\pm0.02$ & $1.40\pm0.31$ & $0.17\pm0.08$ & $1.40\pm0.14$ & $4.5\pm0.3$ & 40.31 & 46.13 & 129.6/106 \\
7204640111 & $0.50$\tnote{f} & $0.67\pm0.05$ & $3.57\pm0.96$ & $2.84\pm0.77$ & $0.23\pm0.07$ & 6.75 & 8.55 & 60.1/45 \\
7204640114 & $0.50$\tnote{f} & $0.79\pm0.05$ & $1.35\pm0.27$ & $2.65\pm0.48$ & $0.15\pm0.03$ & 3.93 & 4.21 & 59.7/66 \\
\bottomrule
\end{tabular}
\begin{tablenotes}
\footnotesize
\item[a] Flux computed in the 0.5--10~keV energy range in units of
$10^{-12}\,\mathrm{erg\,cm^{-2}\,s^{-1}}$.
\item[b] Power-law normalization expressed in units of
$10^{-3}\,\mathrm{photons\,keV^{-1}\,cm^{-2}\,s^{-1}}$ at 1~keV.
\item[f] Parameter fixed during the fit.
\end{tablenotes}
\end{threeparttable}
\end{table*}

To enable comparison with the spectral model adopted for the 2013 outburst \citep{previous_spin}, we additionally fit all four spectra with the alternative model \texttt{tbabs*(diskbb+powerlaw)}, replacing the single-temperature blackbody component with a multicolor disk blackbody; the absorption and power-law components were left identical in form to the \texttt{bbodyrad} fits described above (Table~\ref{tab:specfit}).

For ObsID 7204640103, $N_{\rm H}$ was left free; for ObsID 7204640104, the \texttt{diskbb} fit did not converge with $N_{\rm H}$ free, and $N_{\rm H}$ was likewise fixed at $0.5 \times 10^{22}\ {\rm cm}^{-2}$ for this observation. Uncertainties quoted below are $1\sigma$, from the covariance matrix at the best fit, matching the convention used to report the \texttt{bbodyrad} results in Table~\ref{tab:specfit}.

For the two faintest spectra, ObsIDs 7204640111 and 7204640114 ($N_{\rm H}$ fixed at $0.5 \times 10^{22}\ {\rm cm}^{-2}$ in both models), the \texttt{diskbb} fits return disc temperatures of $kT_{\rm in} = 82 \pm 20\ {\rm eV}$ and $103 \pm 17\ {\rm eV}$, respectively, markedly cooler than typical disc-blackbody temperatures found in accreting neutron-star binaries ($kT_{\rm in} \sim 0.5$--$2\ {\rm keV}$), and lower by an order of magnitude than the corresponding \texttt{bbodyrad} temperatures ($kT = 0.67 \pm 0.05\ {\rm keV}$ and $0.79 \pm 0.05\ {\rm keV}$). The associated \texttt{diskbb} normalizations are essentially unconstrained, with fractional $1\sigma$ uncertainties of $220\%$ (ObsID 7204640111) and $135\%$ (ObsID 7204640114), compared to $27\%$ and $20\%$ for the equivalent \texttt{bbodyrad} normalizations. This is consistent with the well-known temperature--normalization degeneracy in the \texttt{diskbb} model: an arbitrarily large, cool emitting region can reproduce a given soft flux equally well as a smaller, hotter one, and the present count statistics do not break this degeneracy. Unlike the two brighter observations, the power-law index remains close to the canonical range under \texttt{diskbb} for these two spectra ($\Gamma = 1.49 \pm 0.17$ and $1.03 \pm 0.14$), if anything closer to $\Gamma \sim 1$--$2$ than the corresponding \texttt{bbodyrad} values ($\Gamma = 2.84 \pm 0.77$ and $2.65 \pm 0.48$; Table~\ref{tab:specfit}), and the \texttt{diskbb} fits achieve $\chi^2/{\rm dof} = 41.38/43$ and $35.25/64$, comparable to or somewhat better than the \texttt{bbodyrad} values of $60.1/45$ and $59.7/66$.

Taken together, the \texttt{diskbb} model does not provide a uniformly better or worse description of the four spectra than \texttt{bbodyrad}: it achieves comparable or marginally better fit statistics for the two fainter observations, at the cost of an essentially unconstrained normalization and an implausibly low disc temperature, while for the two brighter observations it achieves comparable or marginally worse fit statistics at the cost of a substantially steeper power-law index. In neither regime does \texttt{diskbb} offer a clear improvement over \texttt{bbodyrad}, and in each case its best-fit parameters deviate from the physically expected range for this system in some respect. We therefore retain \texttt{tbabs*(bbodyrad+powerlaw)} as our fiducial model throughout this work, noting that the present NICER count statistics are insufficient to simultaneously and robustly constrain a disc temperature, disc normalization, and independent power-law slope for any of the four spectra.

\subsection{Timing Analysis}
\label{sec3.4}
\begin{figure}[!t]
 \centering
 \includegraphics[width=\columnwidth]{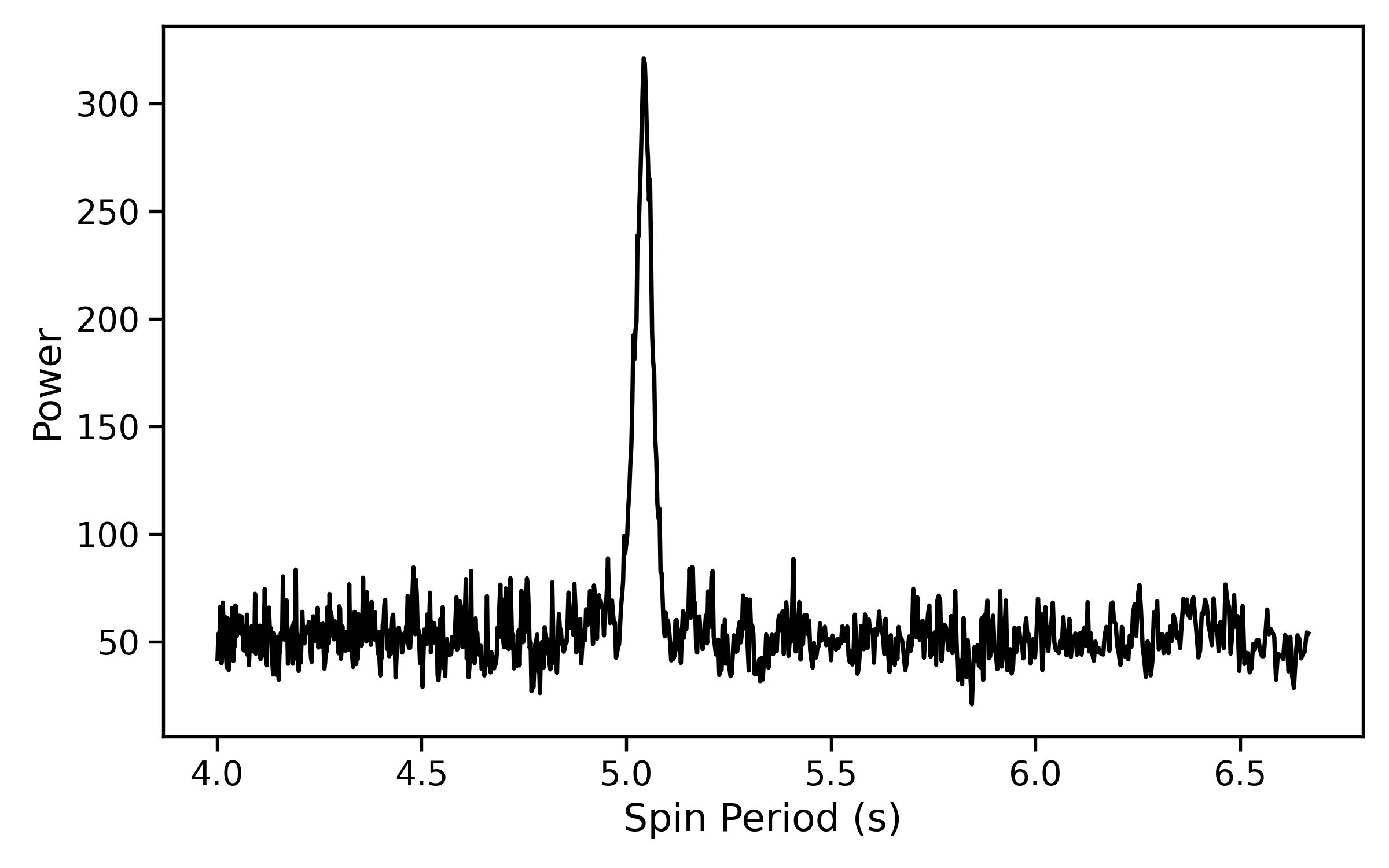}
 \caption{Periodogram of SXP~5.05 from the 2024 NICER observations in the 0.3--10~keV energy range, showing the spin-period peak at 5.04286~s.}
 \label{Figure5}
\end{figure}

In order to understand the pulsation-dependent characteristics of the source, we also perform timing analysis for all observations. (Spin-phase-resolved spectral analysis could not be carried out because of poor statistics.) We use the \texttt{stingray.pulse.search} module from the software \texttt{STINGRAY v2.1}. We calculate the uncertainty in the spin period value by using a bootstrapping approach. We first apply barycentric correction to the event file using \texttt{barycorr}. We filter the event file for an energy range of 0.3--10~keV and search for the period in a frequency band centered on the previously known value of 5.05~s. We observe a significant peak at $5.04286\pm0.00254$~s for ObsID 7204640104, segment 6 (the spin periods of all segments are listed in Table~\ref{tab:gti_timing}), which is slightly lower than the previous value of $5.05006\pm0.00001$~s \citep{previous_spin}.

Pulsation searches were also performed for ObsIDs 7204640111 and 7204640114, the two fainter observations with sufficient counts for spectral fitting, which sample orbital phases $\phi \approx 0.89$ and $\phi \approx 0.14$ under the ephemeris of Eq.~\eqref{eq:phase}, distinct from the $\phi = 0.72$--$0.79$ range of ObsIDs 7204640103 and 7204640104. No statistically significant signal was detected in either observation: the $Z^2_2$ periodograms show no peak near 5.05~s above the noise level, as the source count rates at these epochs are roughly an order of magnitude lower (Table~\ref{tab:specfit}). The present data therefore cannot constrain the orbital-phase dependence of the pulse profile at these phases.

The pulse period search was performed using the $Z^2_n$ statistic \citep{Buccheri1983}, computed directly from the barycenter-corrected photon arrival times, with $n = 2$ harmonics ($Z^2_2$) adopted as the primary detection statistic throughout this work. Since the folded pulse profiles (Figs.~\ref{fig:gti_grid_part1} and \ref{fig:gti_grid_part2}) are predominantly single-peaked, we examined whether this choice of $n$ is justified, or whether a single-harmonic search ($Z^2_1$) is sufficient to describe the signal.

We repeated the period search independently with $n = 1$ and $n = 2$ on an identical frequency grid for each ObsID. For ObsID 7204640104, the $Z^2_2$ search yields a peak statistic of 1172.7 at $P = 5.049571$~s, compared to 1026.7 at the same period for $Z^2_1$, corresponding to an increase of $\Delta Z^2 = 146.0$. This demonstrates that the second harmonic contributes statistically significant signal power rather than merely duplicating the fundamental component. Similarly, for ObsID 7204640103 the peak values are 410.8 ($Z^2_2$) and 373.3 ($Z^2_1$), giving $\Delta Z^2 = 37.5$, again supporting the use of the two-harmonic search.

We also examined whether the binary orbital motion of the neutron star could account for the segment-to-segment period differences in Table~\ref{tab:gti_timing}. The light-travel-time (R{\o}mer) delay introduced by the neutron star's motion about the binary barycentre is given by the Blandford--Teukolsky binary delay model \citep{Blandford1976,Lorimer2004}:
\begin{equation}
\Delta_{\rm R}(t)
=
x \left[
(\cos E - e)\sin\omega
+
\sqrt{1-e^{2}}\,\sin E \cos\omega
\right],
\label{eq:roemer}
\end{equation}
where $x=a_{\rm x}\sin i=142.4$~light-s, $e=0.155$, and $\omega=19.7^{\circ}$ are the projected semimajor axis, eccentricity, and longitude of periastron, respectively, from \citet{previous_spin}. The eccentric anomaly $E$ is the solution of Kepler's equation,
\begin{equation}
E-e\sin E=M,
\label{eq:kepler}
\end{equation}
with the mean anomaly $M=2\pi\phi$, expressed directly in terms of the orbital phase $\phi$ defined in Eq.~\eqref{eq:phase}.

Because $\Delta_{\rm R}(t)$ is the projected light-travel-time delay along the line of sight, its time derivative gives the instantaneous fractional radial velocity of the neutron star \citep{Lorimer2004}:
\begin{equation}
\frac{v(t)}{c}
=
\frac{{\rm d}\Delta_{\rm R}(t)}{{\rm d}t}.
\label{eq:velocity}
\end{equation}
The corresponding observed spin period follows the standard binary-Doppler relation for an accreting X-ray pulsar \citep{Deeter1981,previous_spin}:
\begin{equation}
P_{\rm obs}(t)
=
P_{0}
\left[
1+\frac{v(t)}{c}
\right].
\label{eq:pobs}
\end{equation}
Since $P_{\rm obs}(t)$ is directly proportional to $v(t)/c={\rm d}\Delta_{\rm R}/{\rm d}t$, the rate of change of the apparent spin period over a finite observing window is set by the second derivative of the R{\o}mer delay,
\begin{equation}
\frac{{\rm d}P_{\rm obs}}{{\rm d}t}
=
P_{0}
\frac{{\rm d}^{2}\Delta_{\rm R}(t)}{{\rm d}t^{2}}.
\label{eq:pdot}
\end{equation}
Thus, the apparent spin-period drift is determined by the local curvature, rather than the instantaneous value, of Eq.~\eqref{eq:roemer}. Equations~\eqref{eq:roemer}--\eqref{eq:pdot} therefore provide a direct connection between the orbital elements of \citet{previous_spin} and the predicted spin-period drift, without requiring any additional assumptions beyond the orbital solution already adopted for the phase folding in Eq.~\eqref{eq:phase}.

Evaluating Eq.~\eqref{eq:pobs} numerically across a full orbital cycle, the R{\o}mer delay modulates the observed spin period by up to $\Delta P \sim 3.5$~ms at maximum amplitude, comparable to our smallest per-segment uncertainty of $\sim$2.5~ms (Table~\ref{tab:gti_timing}). However, our coherent pulsation detections are confined to a single $\sim$28~hr window (MJD~60512.73--60513.90), corresponding to orbital phases $\phi=0.72$--$0.79$ via Eq.~\eqref{eq:phase}, and therefore spanning only $\Delta\phi\sim0.07$ of the orbit.

Over this restricted baseline, a linear fit to Eq.~\eqref{eq:pobs}, evaluated at the actual GTI timestamps of Table~\ref{tab:gti_timing}, gives, via Eq.~\eqref{eq:pdot}, an induced spin-period drift of
\begin{equation}
\frac{{\rm d}P_{\rm obs}}{{\rm d}t}
\sim
1.2\times10^{-8}\ {\rm s\,s^{-1}},
\end{equation}
corresponding to a total accumulated shift of only $\sim$1.2~ms across the full observing window. This is smaller than every individual period uncertainty in Table~\ref{tab:gti_timing}, which ranges from approximately 3--24~ms. We therefore conclude that binary Doppler motion cannot account for the scatter observed between segments, nor can it be mistaken for genuine spin evolution over this short baseline. Consequently, we do not apply a per-photon orbital correction given the limited temporal baseline of the coherent pulsation detections.

\begin{figure}[!t]
 \centering
 \includegraphics[width=\columnwidth]{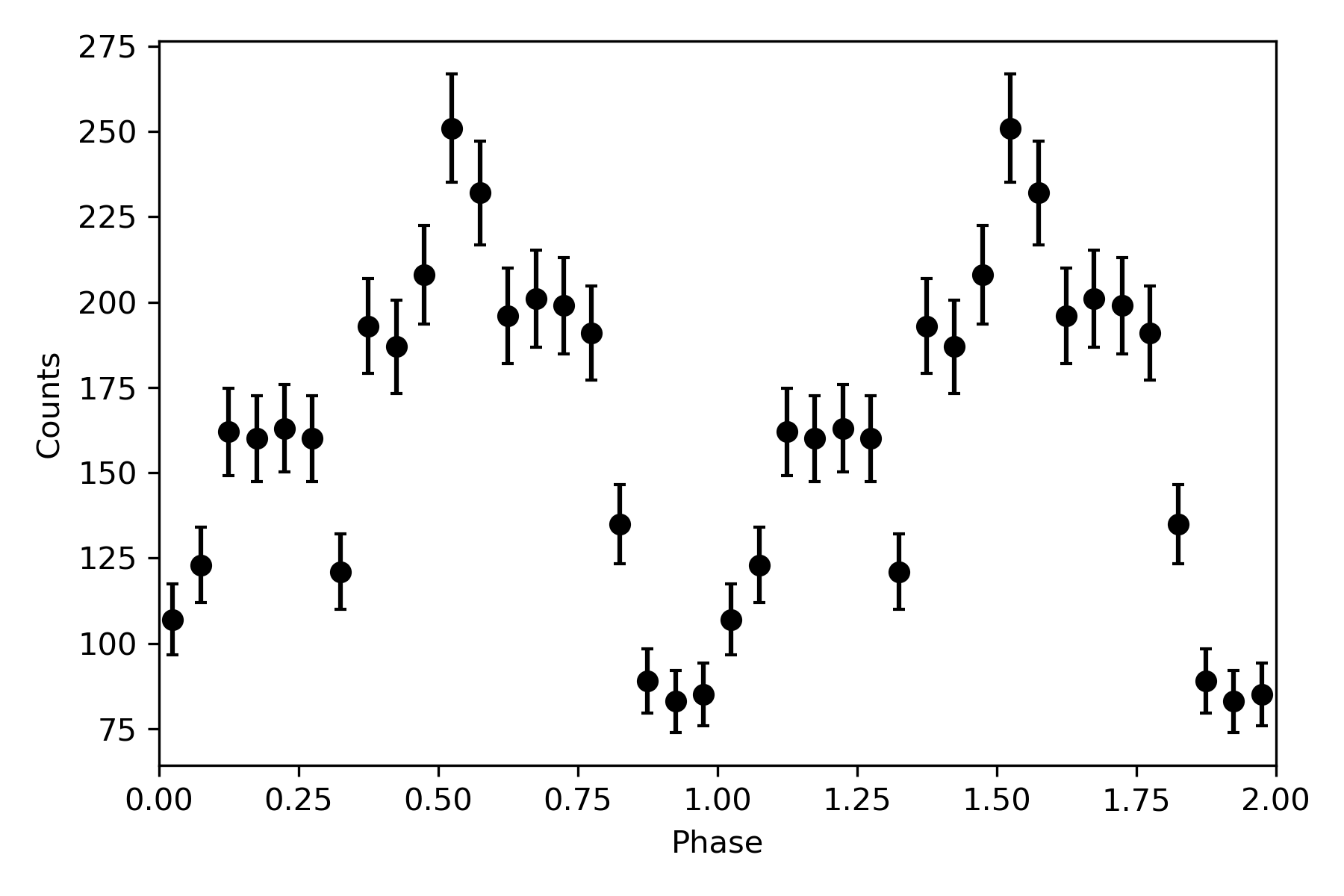}
 \caption{Pulse profile of SXP~5.05 obtained by folding the 0.3--10~keV NICER light curve from the 2024 outburst at $\sim$5.05~s.}
 \label{Figure6}
\end{figure}

\begin{table*}[!t]
\centering
\caption{Spin period of SXP~5.05 for individual GTI segments in the NICER observations where pulsations are detected. Within observations and across them, the spin period remains constant within the error bars.}
\label{tab:gti_timing}
\small
\setlength{\tabcolsep}{9pt}
\renewcommand{\arraystretch}{1.15}
\begin{tabular}{ccccccc}
\toprule
ObsID & GTI & MJD & Period (s) & PF & Amp Ratio & Phase Diff \\
\midrule
7204640103 & 1 & 60512.732107 & $5.041 \pm 0.012$ & $0.54 \pm 0.05$ & $0.31 \pm 0.11$ & $-0.340 \pm 0.037$ \\
7204640103 & 2 & 60512.796554 & $5.046 \pm 0.006$ & $0.66 \pm 0.05$ & $0.40 \pm 0.09$ & $\phantom{-}0.171 \pm 0.020$ \\
7204640103 & 3 & 60512.861078 & $5.043 \pm 0.009$ & $0.57 \pm 0.05$ & $0.36 \pm 0.09$ & $-0.312 \pm 0.028$ \\
7204640104 & 1 & 60513.249286 & $5.043 \pm 0.024$ & $0.58 \pm 0.06$ & $0.29 \pm 0.13$ & $-0.347 \pm 0.046$ \\
7204640104 & 2 & 60513.441832 & $5.042 \pm 0.005$ & $0.67 \pm 0.05$ & $0.33 \pm 0.08$ & $-0.330 \pm 0.021$ \\
7204640104 & 3 & 60513.505359 & $5.053 \pm 0.016$ & $0.62 \pm 0.06$ & $0.55 \pm 0.13$ & $-0.325 \pm 0.026$ \\
7204640104 & 4 & 60513.506844 & $5.056 \pm 0.016$ & $0.64 \pm 0.06$ & $0.62 \pm 0.16$ & $-0.328 \pm 0.026$ \\
7204640104 & 5 & 60513.635440 & $5.049 \pm 0.008$ & $0.56 \pm 0.05$ & $0.28 \pm 0.09$ & $\phantom{-}0.146 \pm 0.023$ \\
7204640104 & 6 & 60513.701906 & $5.044 \pm 0.003$ & $0.53 \pm 0.04$ & $0.50 \pm 0.07$ & $-0.373 \pm 0.018$ \\
7204640104 & 7 & 60513.831359 & $5.047 \pm 0.004$ & $0.59 \pm 0.05$ & $0.62 \pm 0.10$ & $\phantom{-}0.153 \pm 0.020$ \\
7204640104 & 8 & 60513.895864 & $5.043 \pm 0.006$ & $0.49 \pm 0.04$ & $0.50 \pm 0.10$ & $\phantom{-}0.200 \pm 0.021$ \\
\bottomrule
\end{tabular}
\end{table*}

\begin{figure}[!t]
 \centering
 \includegraphics[width=\columnwidth]{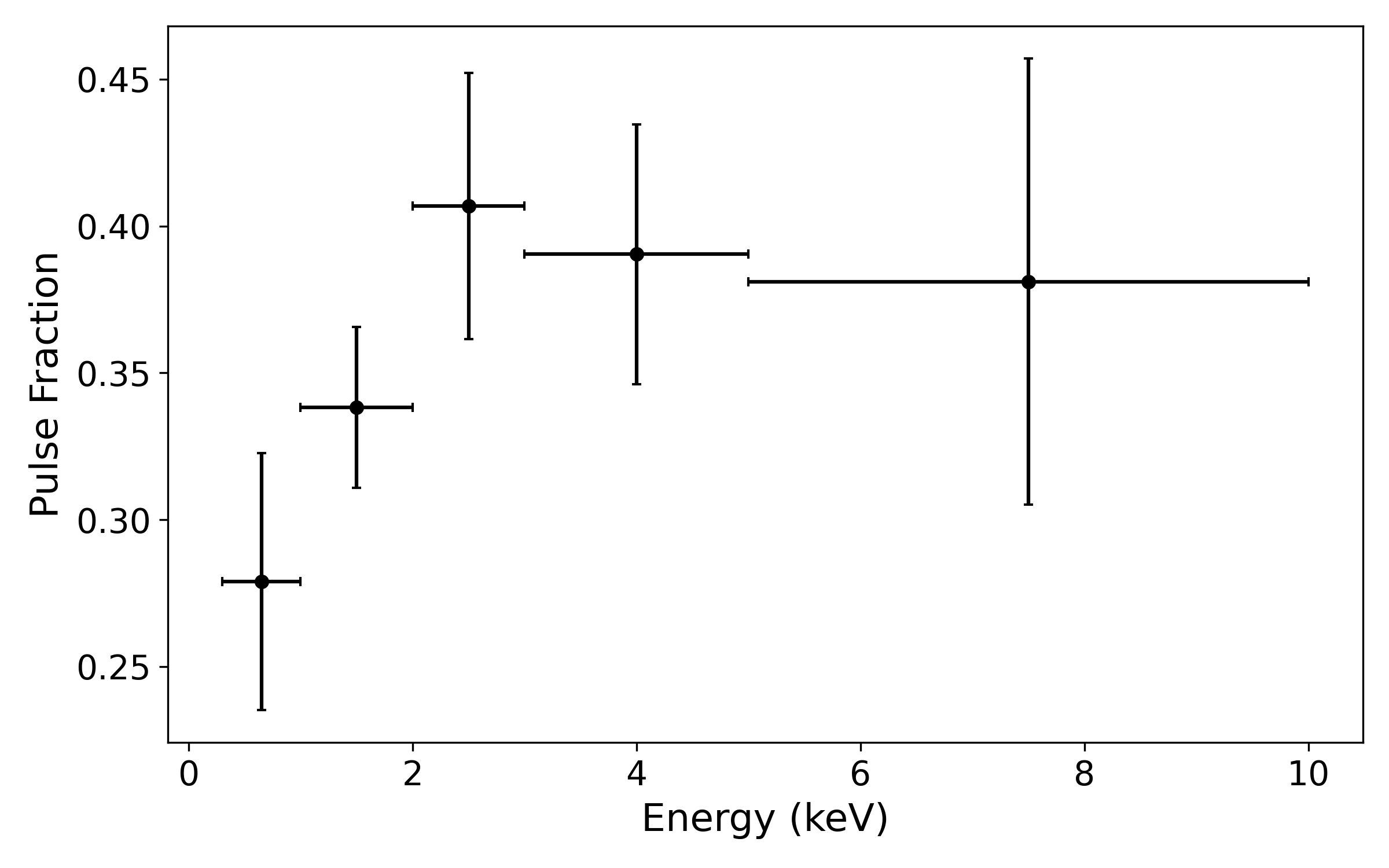}
 \caption{Variation of the pulse fraction with energy for SXP~5.05 during its 2024 outburst observed with NICER. The trend is characteristic of HMXB pulsars, where the pulse fraction increases with energy until it reaches a cut-off.}
 \label{Figure7}
\end{figure}

\subsection{OGLE Timing}

\begin{figure*}[!t]
    \centering
    \includegraphics[width=0.95\textwidth]{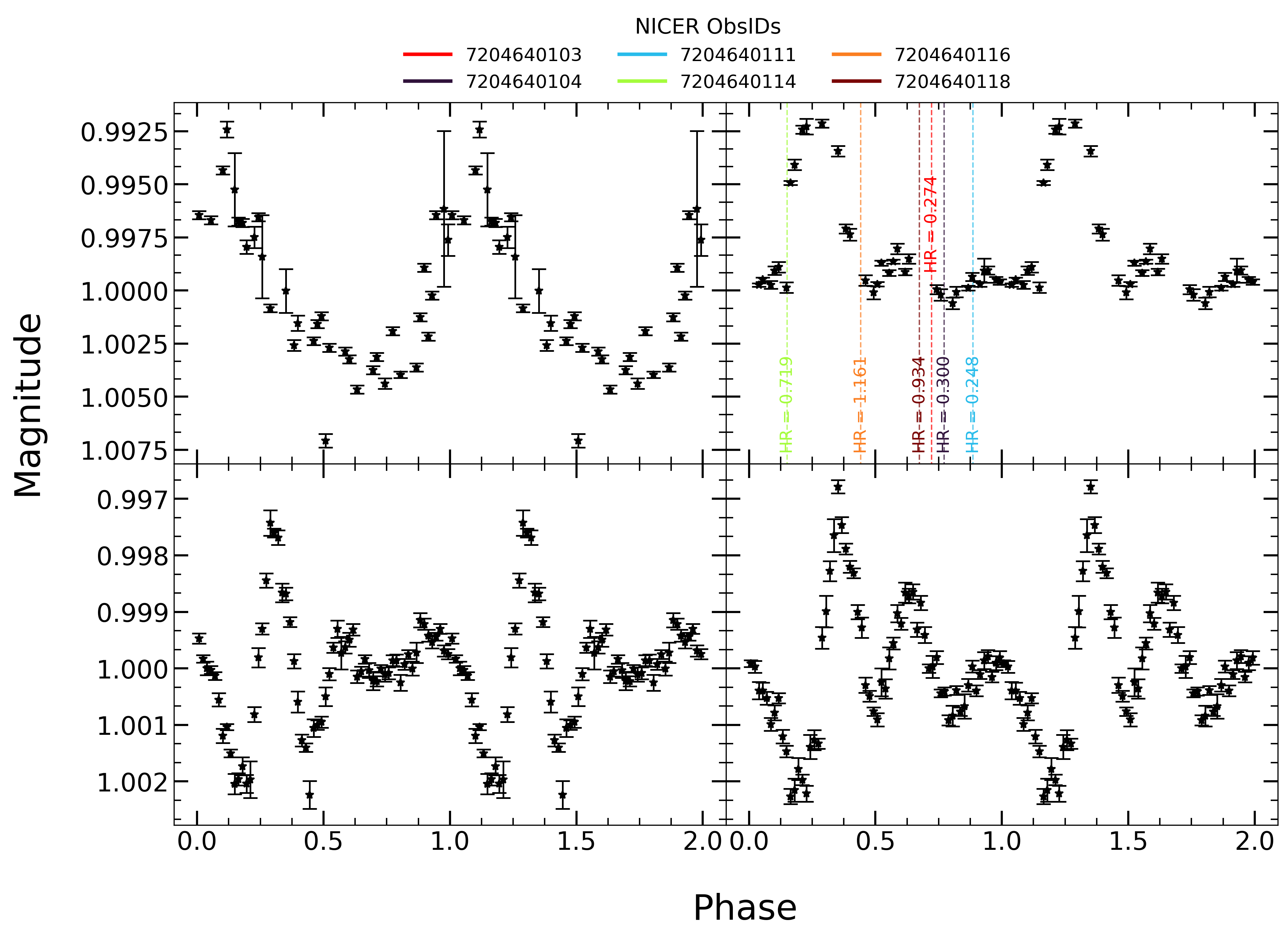}
    \caption{\textit{I}-band light curves of SXP~5.05 phase-folded on the orbital period $P_{\mathrm{orb}} = 17.13$~d, with reference epoch $T_{0} = 56680.45$~MJD (Table~6 of \citealt{previous_spin}), plotted over two consecutive cycles for clarity. \textbf{Top left:} Outburst epoch~1 (MJD 56575--56707; $\sim$132~d). \textbf{Top right:} Outburst epoch~2 (MJD 60455--60565; $\sim$110~d); the dashed vertical lines mark the orbital phases of the six contemporaneous \textit{NICER} X-ray observations shown in Fig.~\ref{fig:Figure1} (ObsIDs 7204640103, 7204640104, 7204640111, 7204640114, 7204640116, and 7204640118), color-coded as indicated in the legend; each line is labeled with the corresponding X-ray hardness ratio (HR, as defined in Section~\ref{subsec:HID}) in the same color as its ObsID. \textbf{Bottom left:} Quiescent phase preceding the 2024 outburst (MJD 60100--60350; $\sim$250~d). \textbf{Bottom right:} Quiescent phase following the 2024 outburst (MJD 60550--60730; $\sim$180~d).}
    \label{fig:optical_phase_panels}
\end{figure*}

Figure~\ref{fig:optical_phase_panels} presents the \textit{I}-band light curves of SXP~5.05 phase-folded at $P_\mathrm{orb} = 17.13$~d across four temporal intervals. The two outburst profiles (top panels) differ substantially in both amplitude and morphology. Outburst epoch~1 (top left; MJD 56575--56707) shows the largest amplitude modulation, with a well-defined asymmetric profile. Outburst epoch~2 (top right; MJD 60455--60565) presents a markedly different morphology: the modulation is weaker in amplitude, the profile is broader and structurally less organised, and the sharp asymmetric ingress prominent in epoch~1 is absent. The flux distribution across orbital phase in epoch~2 lacks a single well-defined minimum, indicating a qualitatively different pattern of variability compared to epoch~1. The top-right panel additionally marks the orbital phases of six contemporaneous X-ray observations with colored dashed lines; the corresponding hardness ratios are indicated next to each line, using the same color as the associated ObsID (in the range $\mathrm{HR} \approx 0.25$--$1.16$). This allows a direct comparison of the optical state of the source with its X-ray spectral hardness at each epoch.

The bottom panels show the quiescent intervals immediately preceding (left; MJD~60100--60350, $\sim$250~d, $\sim$14.6 orbital cycles) and following (right; MJD~60550--60730, $\sim$180~d, $\sim$10.5 orbital cycles) the 2024 outburst. Both exhibit clear, measurable orbital modulation, though at substantially reduced amplitude relative to either outburst epoch. The pre-outburst profile shows structured low-level variability distributed across the full phase range, with a broad modulation pattern that is distinct in shape from either outburst profile.

The post-outburst profile likewise shows clear variability but with a notably different morphology from the pre-outburst state: in particular, a more pronounced rise-and-dip feature is present near $\phi \approx 0.6$--$0.9$, which is reproduced coherently across both displayed cycles. This feature is comparatively weaker or absent at the same phase range in the pre-outburst profile, indicating that the orbital modulation pattern underwent a structural change between the two quiescent intervals bracketing the 2024 outburst. Across all four panels, the profiles are non-sinusoidal, and no two epochs share an identical morphology, highlighting the epoch-to-epoch variability of the orbital light-curve shape in this system.

\section{Discussion}\label{sec:discussion}
\subsection{Outburst Duration and Fluence}
The 2024 outburst of SXP~5.05, as observed with NICER, spans approximately MJD~$\sim$60512 to $\sim$60546, corresponding to a duration of $\sim$34 days based on the available X-ray coverage. Within this interval, the source exhibits a monotonic decline in count rate from $\sim$8--10~counts~s$^{-1}$ to $\sim$1~counts~s$^{-1}$ (Figure~\ref{fig:Figure1}), indicating that the observations likely sample the decay phase of the outburst rather than its onset.

In contrast, the 2013 outburst reported by \citet{previous_spin} was a significantly longer and more luminous event, with X-ray activity persisting over a substantially extended duration and reaching higher peak luminosities. Although the NICER observations do not fully capture the rise phase of the 2024 outburst, the shorter observed duration and lower peak count rate suggest that the 2024 event is intrinsically weaker than the 2013 outburst.

A useful way to quantify this difference is through the outburst fluence, defined as the time-integrated flux. While a precise fluence estimate would require complete coverage of the outburst, a relative comparison can still be made. The lower count rates and shorter duration of the 2024 event imply a significantly reduced fluence compared to the 2013 outburst. This is consistent with the optical behavior of the system, where the OGLE light curves show a smaller amplitude during the 2024 epoch ($\Delta I \sim 0.15$~mag) compared to the $\sim$0.30~mag variation observed during the 2013 event (Fig.~\ref{fig:optical_outburst_lc}).

The reduced optical amplitude suggests that the circumstellar disc of the Be companion was less extended or less dense during the 2024 outburst. Since the X-ray luminosity in Be/X-ray binaries is directly linked to the amount of material transferred from the disc to the neutron star, a smaller disc naturally leads to a lower accretion rate and hence a reduced X-ray fluence.

Therefore, the combined X-ray and optical evidence indicates that the 2024 outburst represents a comparatively weaker accretion episode, likely driven by a less developed circumstellar disc. This difference in disc state between the two epochs may also have important implications for the observed X-ray--optical properties.

\subsection{Evolution of Optical Orbital Profiles}

The phase-folded optical light curves of SXP~5.05 (Figure~\ref{fig:optical_phase_panels}) reveal significant variability both across different outburst epochs and within individual epochs. A particularly notable feature is the presence of a recurrent dip/eclipse-like structure around orbital phase $\sim$0.7--0.8, which is most clearly seen during the 2013 outburst and remains detectable, albeit weaker, in the 2024 data.

The persistence of this feature at a fixed orbital phase even after 10 years strongly indicates that it is not due to stochastic variability, but instead reflects a stable geometric or density structure within the system. In the context of Be/X-ray binaries, such phase-locked features are commonly associated with asymmetries in the circumstellar disc, such as enhanced density regions, warps, or interaction zones where the neutron star perturbs the disc material during its orbit.

Comparing the two epochs, the 2013 optical orbital profile exhibits stronger modulation and a more pronounced dip near phase $\sim$0.75, consistent with a larger and denser circumstellar disc. In contrast, the 2024 profiles show a reduced amplitude and a less distinct dip, indicating a comparatively weaker or less stable disc. This is consistent with the smaller optical variability amplitude observed during the 2024 outburst ($\Delta I \sim 0.15$~mag) compared to the $\sim$0.30~mag variation in 2013.

Further evolution is observed within the 2024 epoch itself. The pre-outburst profile displays relatively modest modulation, while during the outburst the orbital structure becomes more pronounced, including the emergence of a clearer dip feature. In the post-outburst phase, the modulation weakens again and the profile becomes more irregular, suggesting a reconfiguration or partial dissipation of the disc. This temporal evolution implies that the disc undergoes significant structural changes on timescales comparable to the outburst duration.

The phase location of the dip near $\sim$0.7--0.8 is particularly interesting in the context of the system geometry. Given that SXP~5.05 is known to exhibit X-ray eclipses and strong absorption events, this optical feature may trace regions of enhanced density or vertical structure in the disc that also contribute to X-ray obscuration. In this scenario, the neutron star periodically interacts with or passes behind these denser regions, leading to correlated signatures in both optical and X-ray bands.

The persistence of the dip-like feature near orbital phase $\sim$0.7--0.8 across multiple epochs is also noteworthy in the context of the warped-disc interpretation proposed by \citet{Brown2019}. Their modelling of the 2013 outburst demonstrated that the complex X-ray variability of SXP~5.05 could be explained by a non-axisymmetric circumstellar disc with significant vertical structure. The continued presence of an optical dip at a similar orbital phase in the 2024 data suggests that comparable large-scale disc asymmetries may persist over timescales of years, despite substantial changes in the overall disc strength and outburst properties.

Overall, the optical orbital profiles provide evidence for a non-axisymmetric and time-evolving circumstellar disc. The differences between the 2013 and 2024 epochs, together with the intra-epoch evolution observed in 2024, indicate that each outburst is associated with a distinct disc configuration. When combined with the X-ray spectral and timing behavior, these results reinforce the interpretation of SXP~5.05 as a system in which the neutron star probes the dynamic structure of the Be star disc.

\subsection{Disc Size Estimates from Optical Variability}

The optical variability of Be stars is primarily driven by emission from the circumstellar disc, and can therefore be used as a proxy for changes in disc size and density.

We observe that the 2013 outburst showed $\Delta I \sim 0.3$~mag, whereas the 2024 outburst only showed $\Delta I \sim 0.15$~mag. This suggests that the circumstellar disc during the 2024 outburst was moderately smaller than during the 2013 event. However, this estimate should be treated with caution, as the optical flux also depends on the disc density distribution, inclination, and temperature. Therefore, the observed difference in optical amplitude should not be used quantitatively to measure the disc radius; rather, it qualitatively reflects variations in disc density and structure.

\subsection{Pulse Period and Pulse Profile Properties}

Coherent pulsations are clearly detected in the NICER observations of SXP~5.05, with a measured spin period of $P = 5.04286 \pm 0.00254$~s (ObsID 7204640104, segment 6). This value is consistent, within uncertainties, with previously reported measurements for the system \citep[e.g.,][]{previous_spin}, indicating no statistically significant deviation from the known spin period.

As shown in Section~\ref{sec3.4}, the expected binary Doppler modulation over the $\sim$28~hr baseline of our pulsation detections ($\sim$1.2~ms) is smaller than our individual period measurement uncertainties (Table~\ref{tab:gti_timing}), and we therefore do not attempt to infer short-term spin evolution or orbital modulation from the segment-to-segment period differences. The observed period is interpreted as being consistent with the long-term spin behavior of the source.

The pulse fractions measured for individual GTI segments are listed in Table~\ref{tab:gti_timing} and span a relatively narrow range of $\sim$0.49--0.67, with a mean value of $\sim$0.59. This indicates that the pulsed emission remains a dominant component of the observed X-ray flux throughout the monitored portion of the outburst \citep{kaushik2025spin}. No systematic evolution of pulse fraction with time is evident within the statistical uncertainties, suggesting that the overall beaming geometry and accretion configuration remain broadly stable during the observations.

The pulse profiles show clear periodic modulation and are detected across the observed energy range (see Appendix~\ref{sec:appendix}). The shape of the pulse profile is broadly stable, though variations in amplitude and structure are observed between different observations \citep{cappallo2020geometry}. In particular, the pulse fraction is found to vary with both time and energy, with indications that the modulation is generally stronger at higher energies \citep{kumar2025variability}. The energy-dependent pulse fraction shown in Figure~\ref{Figure7} increases from approximately 0.3--0.4 at lower energies to values approaching 0.8 at higher energies, although with larger uncertainties in the highest-energy bins. Such behaviour is commonly observed in accretion-powered X-ray pulsars and is generally attributed to the increasing dominance of magnetically channelled emission at higher energies \citep{tobrej2025spectro}.

In some observations, the pulsations appear weaker or less well defined. This may be attributed to a combination of reduced count rate and increased scattering or absorption, particularly during phases where the source is partially obscured. However, given the limited signal-to-noise in these intervals, no strong conclusions are drawn regarding changes in the emission geometry.

Overall, the timing analysis confirms the persistent pulsar nature of SXP~5.05, with pulse properties that are consistent with previous studies and typical of accretion-powered X-ray pulsars \citep{roy2022modeling}. The observed variations in pulse fraction and profile shape are likely linked to changes in viewing geometry and local absorption, but a detailed phase-resolved analysis would be required to quantify these effects.

\section{Conclusion}\label{sec:conc}
In this work, we have investigated the X-ray spectral and timing evolution of SXP~5.05 during its 2024 outburst using NICER observations, complemented by optical light curves from OGLE. We have further compared the properties of the 2024 outburst with those observed during the 2013 event to understand the long-term evolution of the system.

Our main findings are summarized as follows:

\begin{itemize}
\item The 2024 outburst is shorter and fainter than the 2013 event, with a lower peak count rate and reduced overall fluence, indicating a comparatively weaker accretion episode.

\item The hardness evolution shows a transition from a soft, high-intensity state to a harder, low-intensity state.

\item Optical observations reveal a smaller variability amplitude during the 2024 outburst compared to 2013, suggesting that the circumstellar disc was less extended or less dense.

\item The orbital-phase-folded optical light curves exhibit a phase-locked dip-like feature around $\phi\sim$0.7--0.8, which persists across epochs but varies in strength. This indicates the presence of a stable, non-axisymmetric structure in the disc, likely associated with enhanced density or disc asymmetry.

\item Significant evolution in the optical orbital profiles is observed both between the 2013 and 2024 epochs and within the 2024 epoch itself (pre-, during, and post-outburst), implying that the disc undergoes substantial structural changes on relatively short timescales.

\item Coherent pulsations at $\sim$5.05~s are detected throughout the observations, with the measured period consistent with previous studies within uncertainties. Variations in pulse fraction and profile shape are observed but are not sufficient to infer detailed changes in emission geometry.

\item The combined X-ray and optical results support a scenario in which SXP~5.05 behaves as an eclipsing Be/X-ray binary where the neutron star probes the structure of the circumstellar disc. The observed variability is primarily governed by changes in disc density, geometry, and viewing conditions.
\end{itemize}

Overall, our results highlight the importance of coordinated X-ray and optical observations in understanding the dynamic behaviour of Be/X-ray binaries. Future observations with improved orbital phase coverage and phase-resolved spectroscopy will be crucial for constraining the disc structure and its interaction with the neutron star in greater detail.

\section*{Acknowledgements}
This research has made use of software provided by the High Energy Astrophysics Science Archive Research Center (HEASARC), which is a service of the Astrophysics Science Division at NASA/GSFC.

\bibliographystyle{sxp505}
\bibliography{sample7}


\appendix
\section{Pulse Profiles and Pulse-Fraction Spectra of Individual GTI Segments}\label{sec:appendix}

In this appendix, we provide the power spectra, pulse profiles, and pulse-fraction spectra for the individual GTI segments of ObsIDs 7204640103 and 7204640104 (Figs.~\ref{fig:gti_grid_part1}--\ref{fig:pf_energy_all}).

\begin{figure*}[!htbp]
\centering
\setlength{\tabcolsep}{4pt}
\begin{tabular}{cc}
\includegraphics[width=0.34\textwidth]{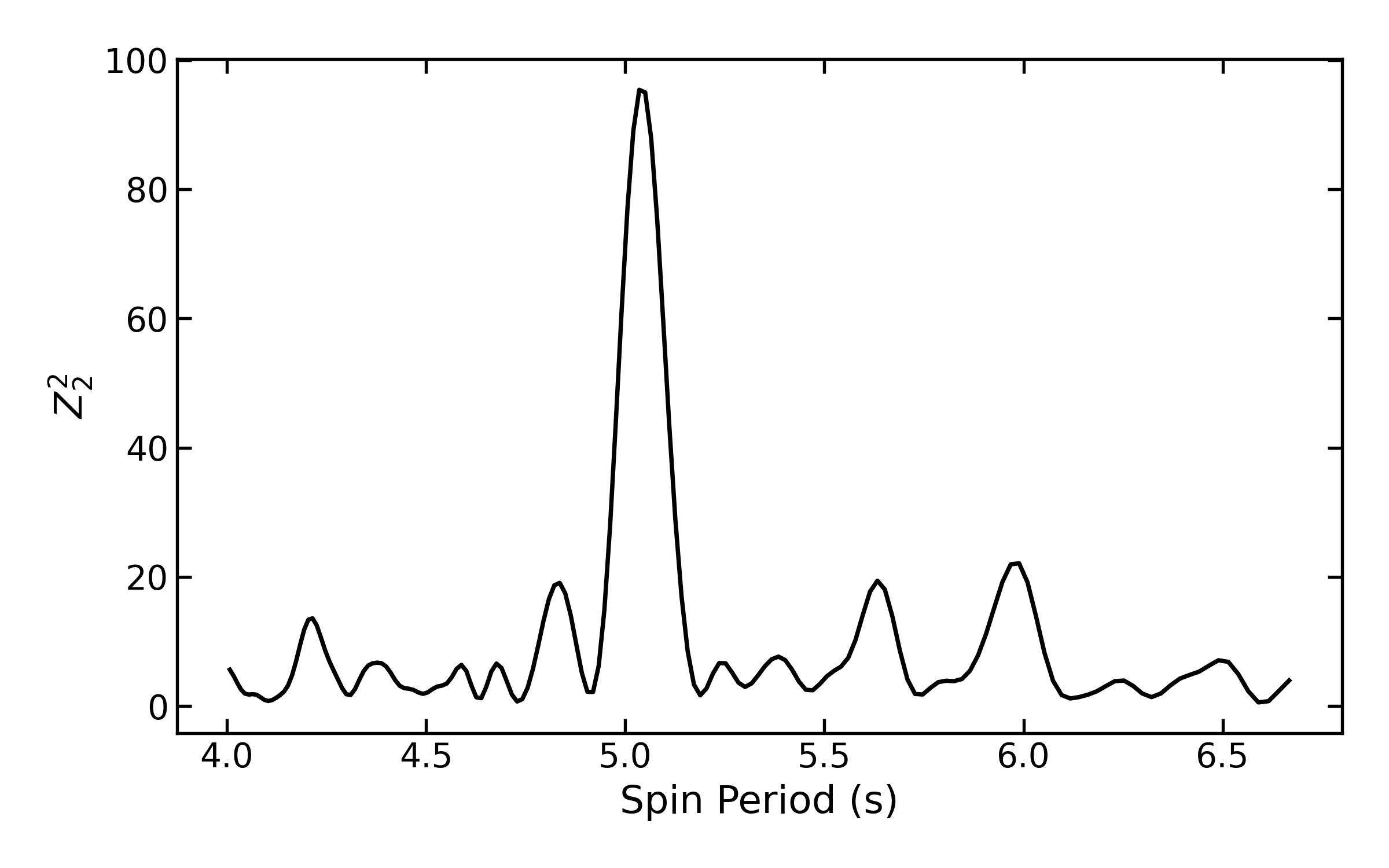} &
\includegraphics[width=0.34\textwidth]{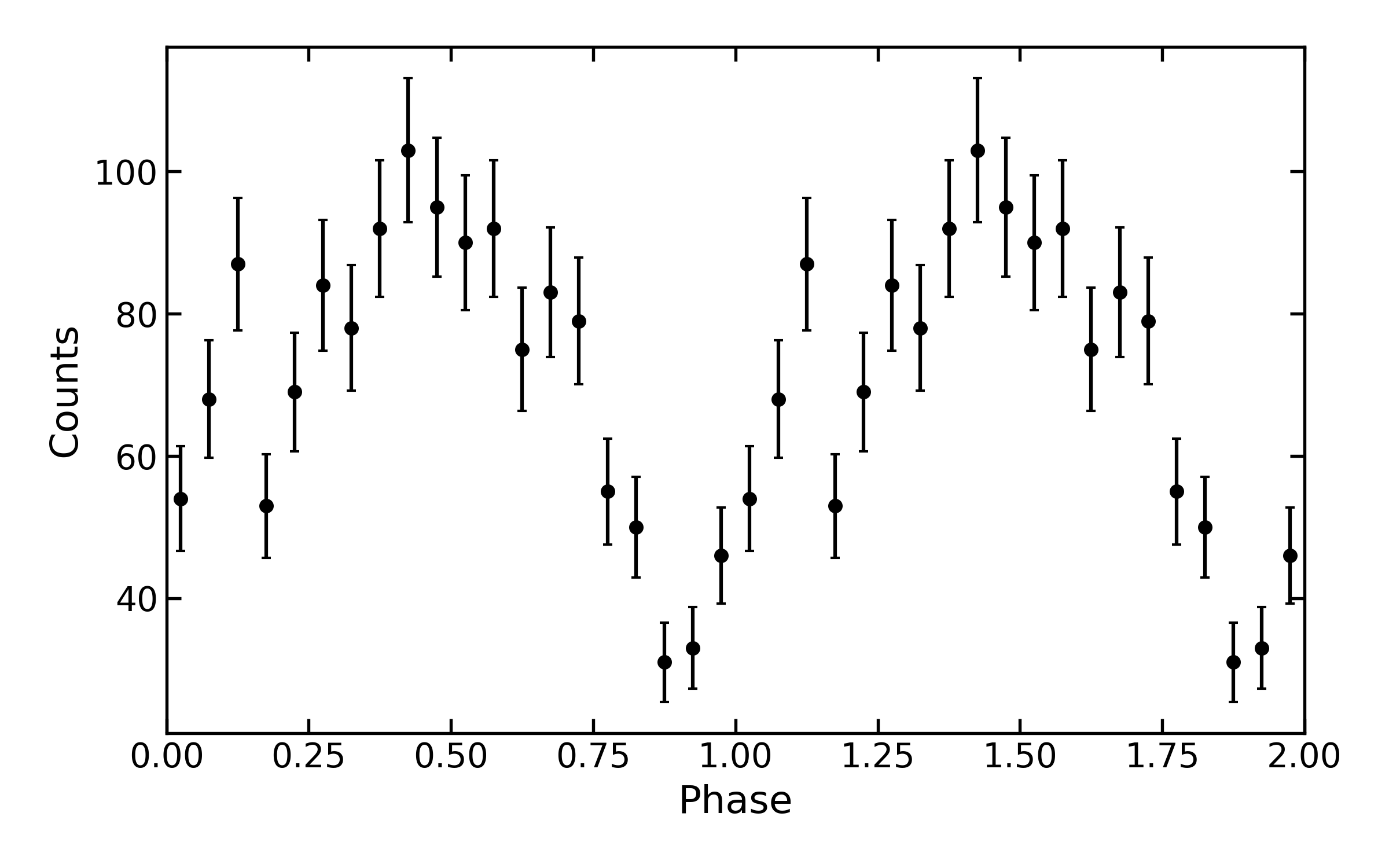} \\
\includegraphics[width=0.34\textwidth]{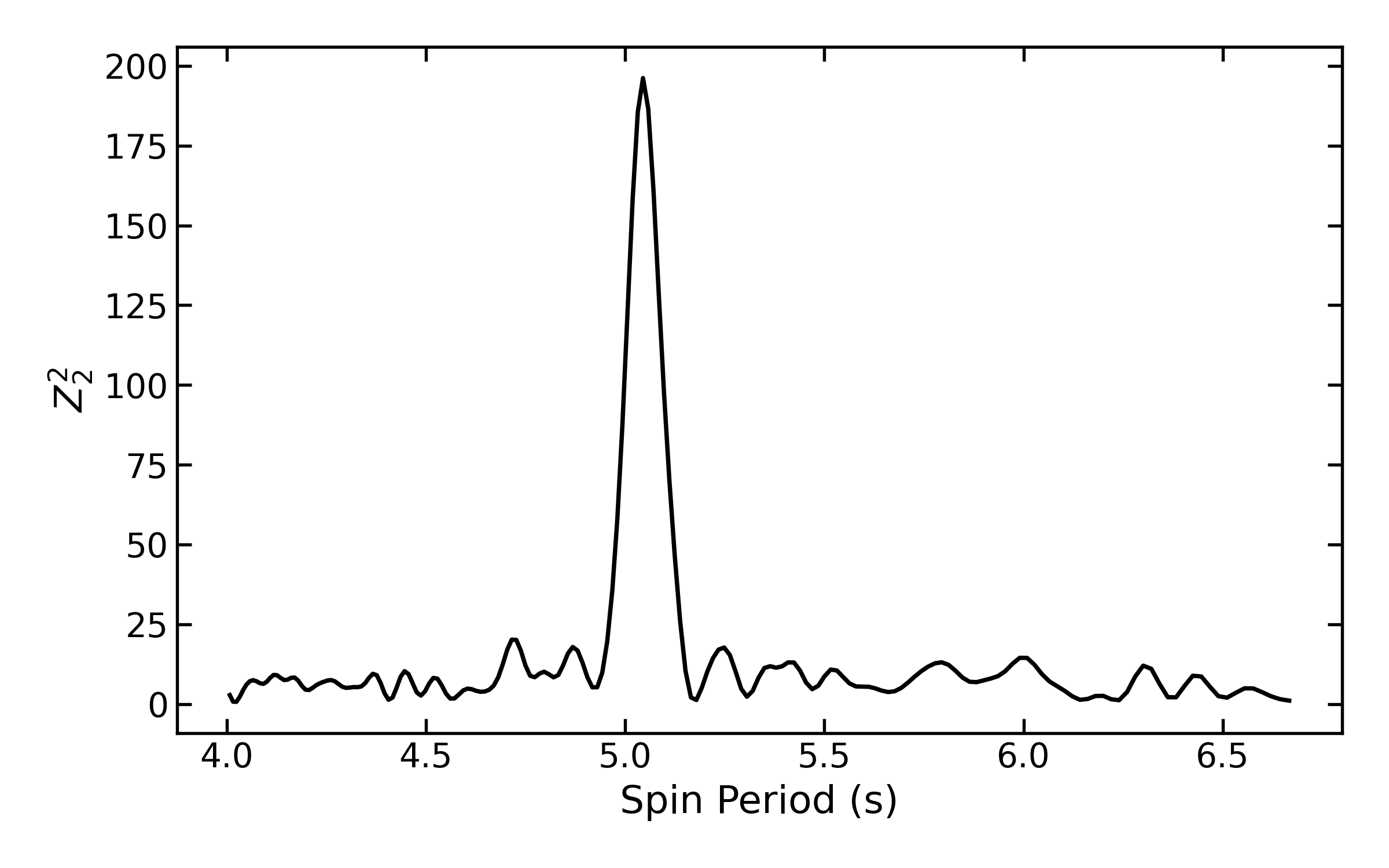} &
\includegraphics[width=0.34\textwidth]{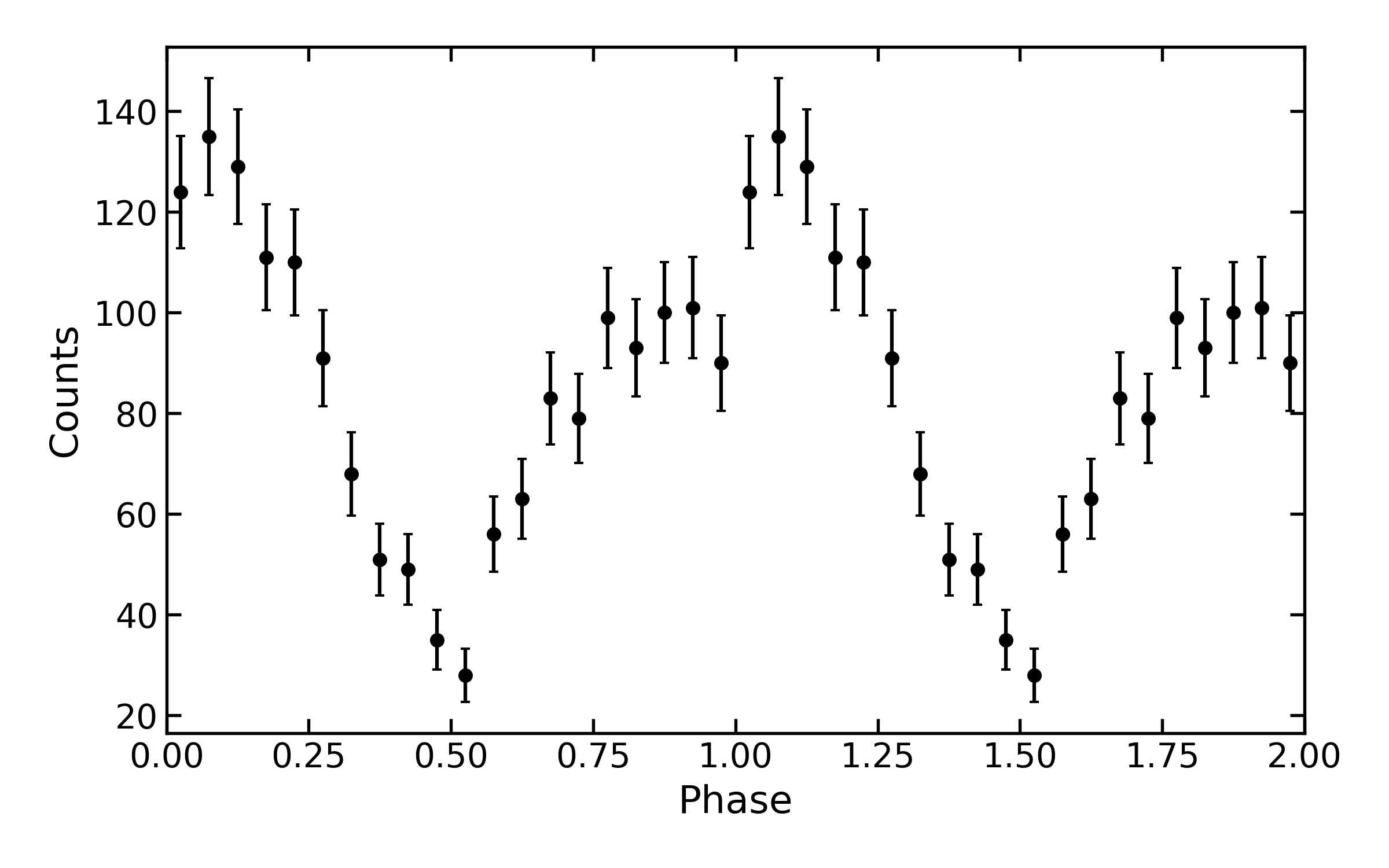} \\
\includegraphics[width=0.34\textwidth]{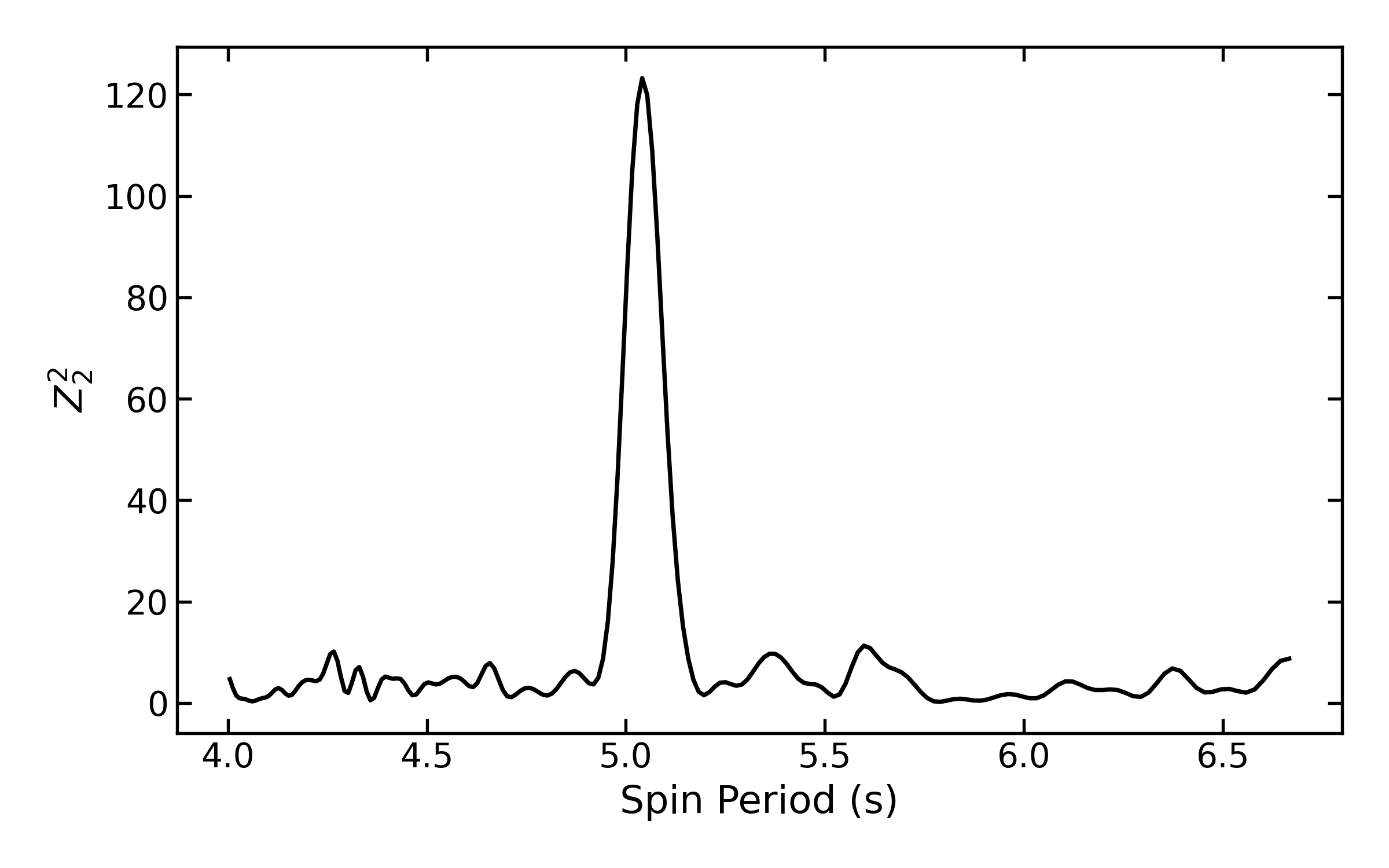} &
\includegraphics[width=0.34\textwidth]{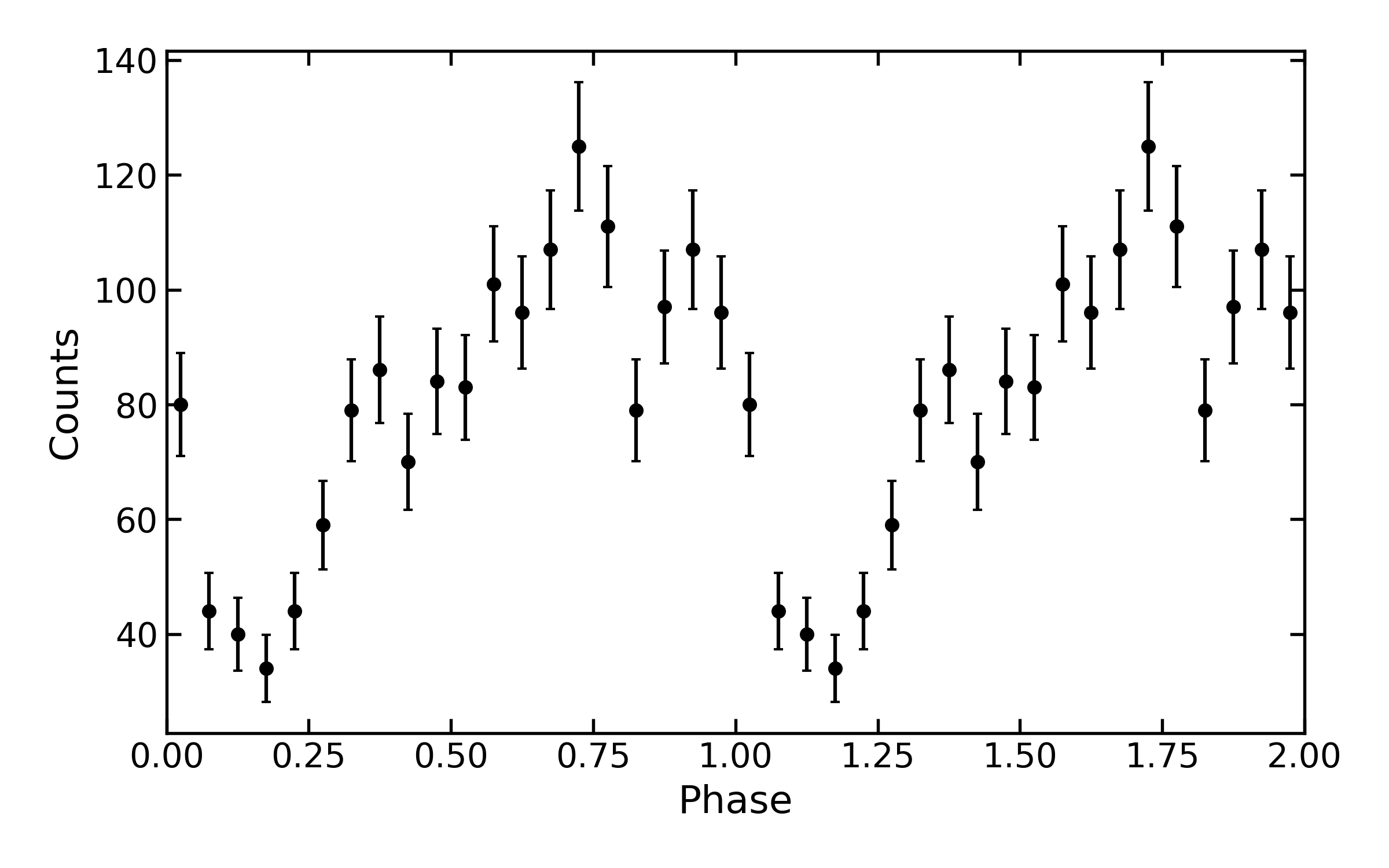} \\
\includegraphics[width=0.34\textwidth]{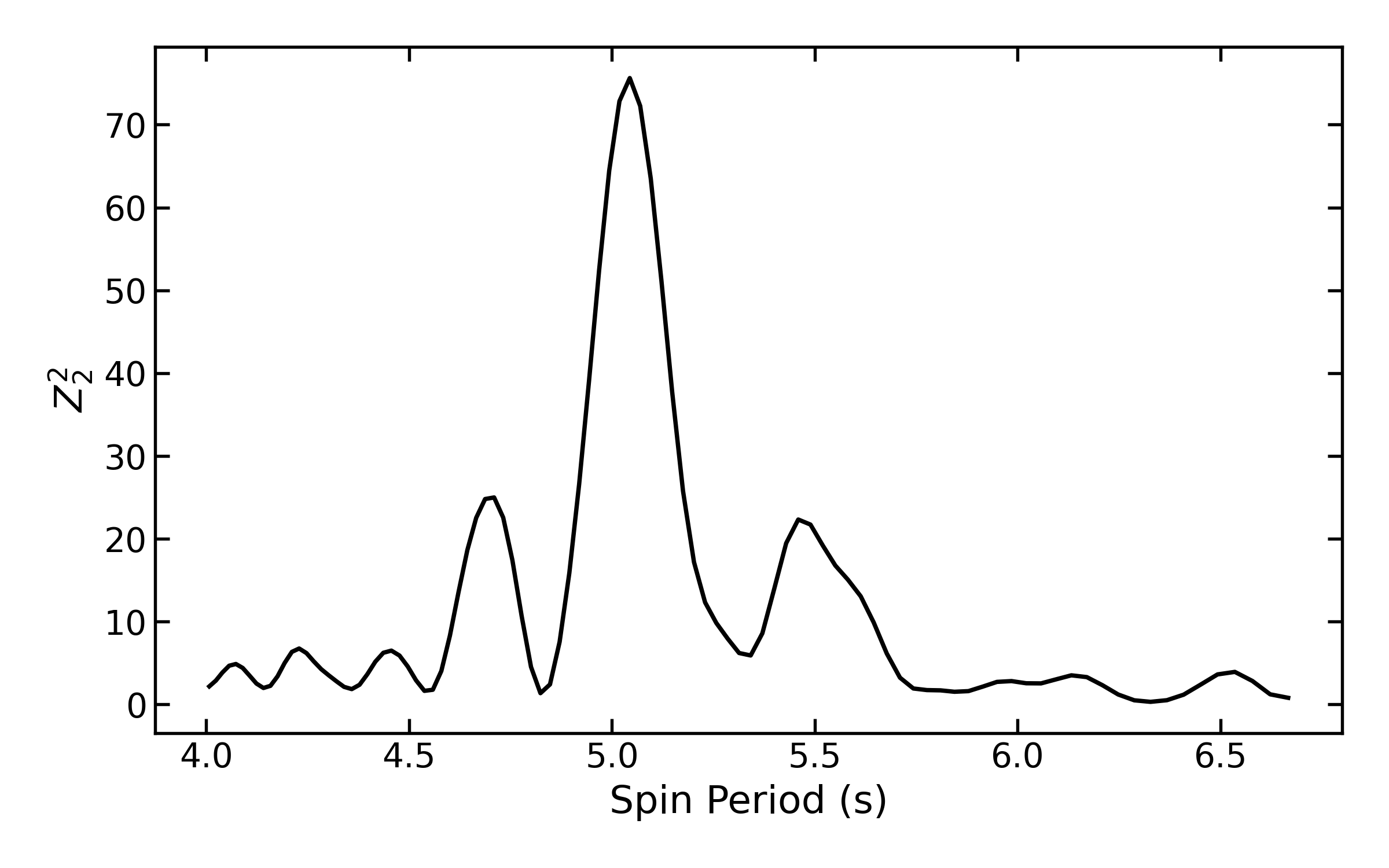} &
\includegraphics[width=0.34\textwidth]{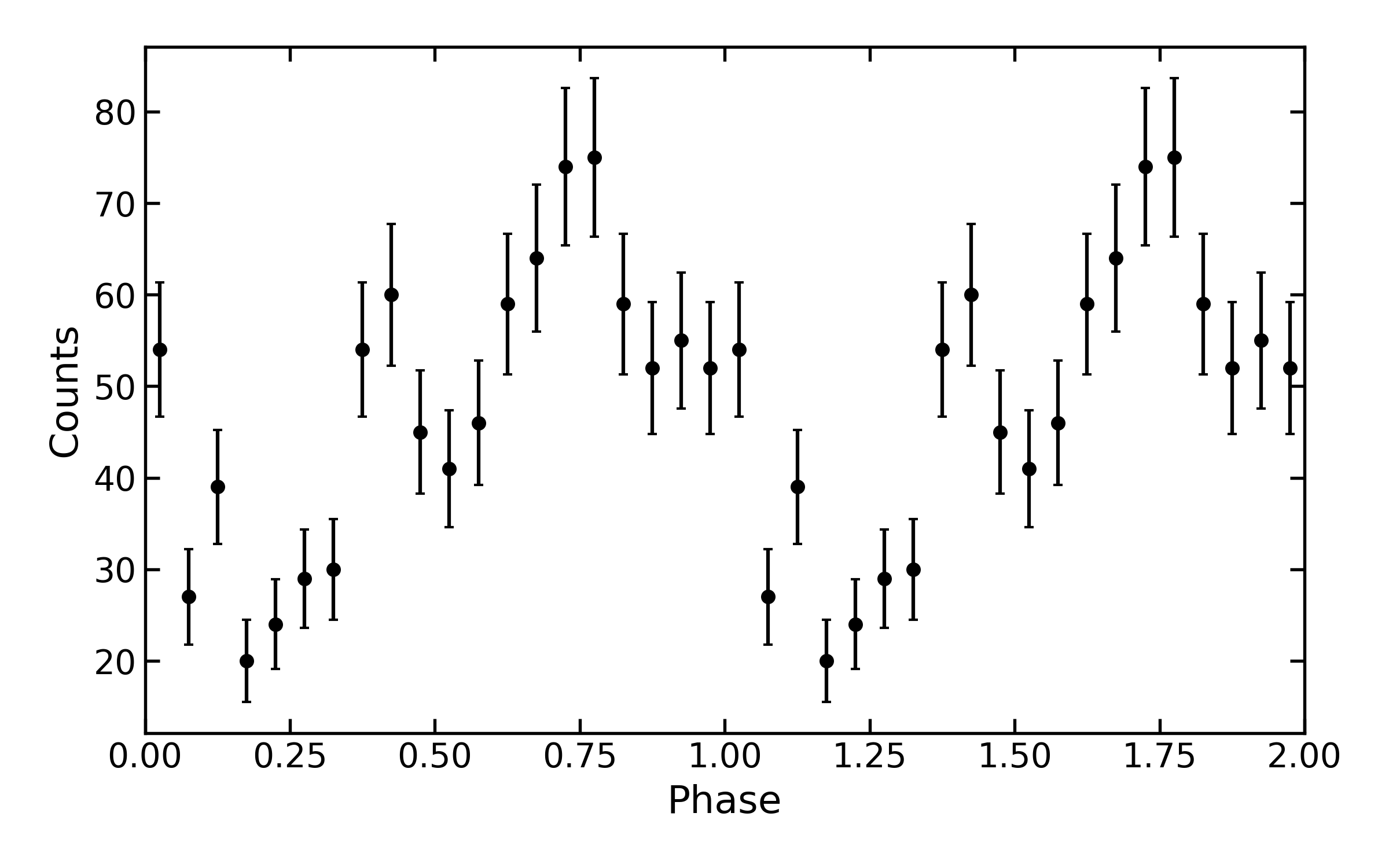} \\
\includegraphics[width=0.34\textwidth]{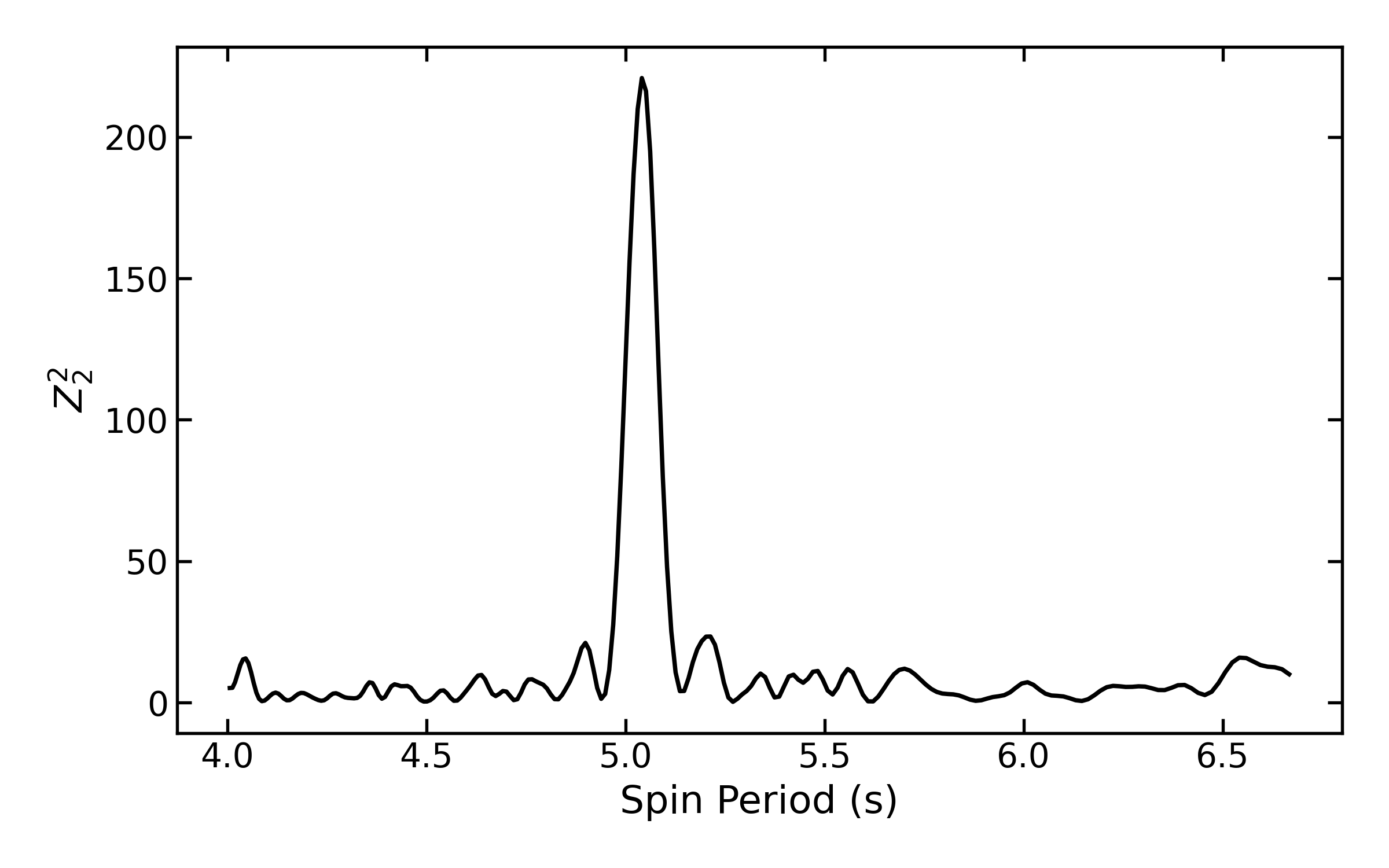} &
\includegraphics[width=0.34\textwidth]{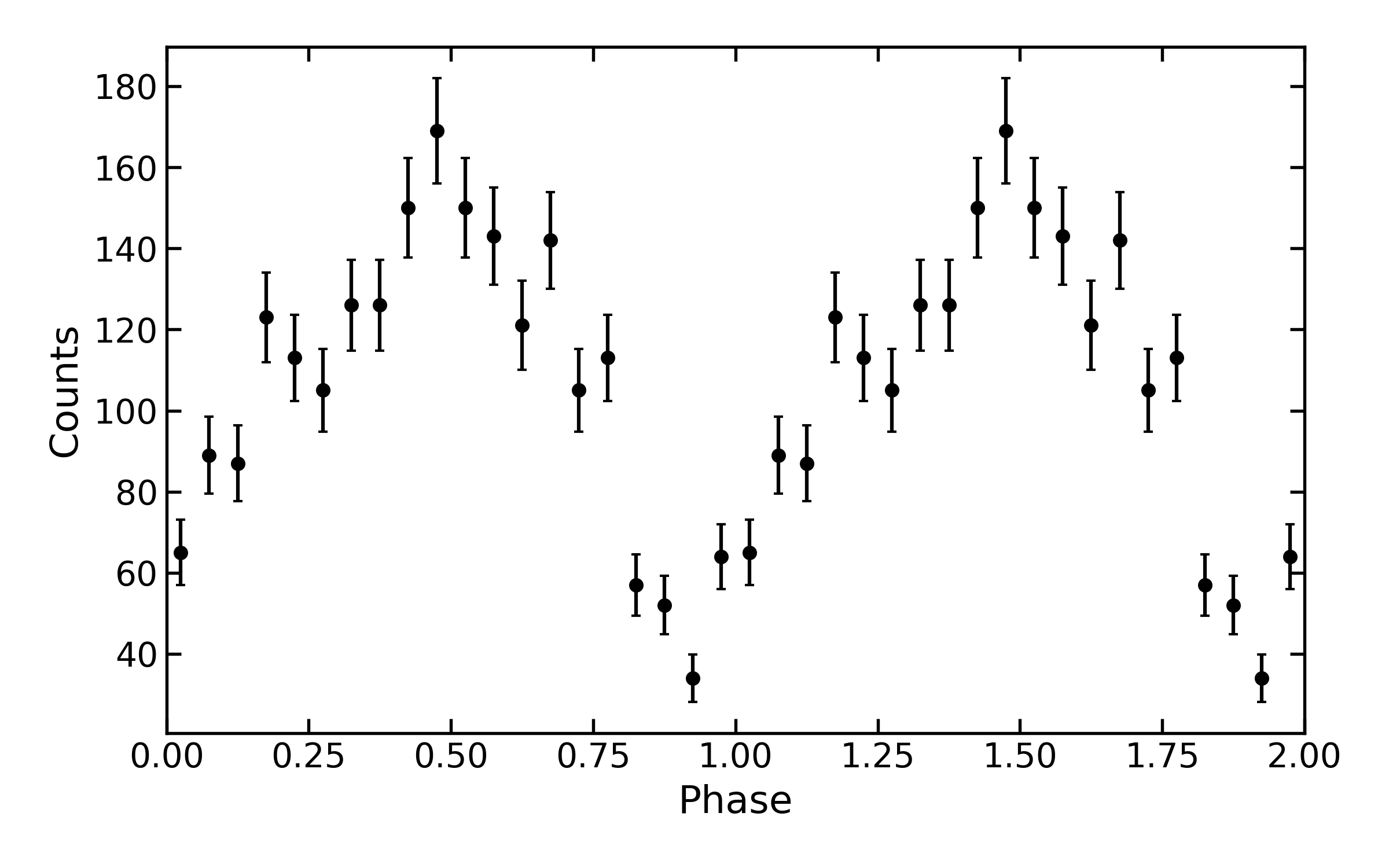} \\
\end{tabular}
\caption{Power spectra (left) and pulse profiles (right) for individual GTI segments. Rows from top to bottom: ObsID 7204640103, GTI segments 1--3; ObsID 7204640104, GTI segments 1--2.}
\label{fig:gti_grid_part1}
\end{figure*}

\begin{figure*}[!htbp]
\centering
\setlength{\tabcolsep}{4pt}
\begin{tabular}{cc}
\includegraphics[width=0.33\textwidth]{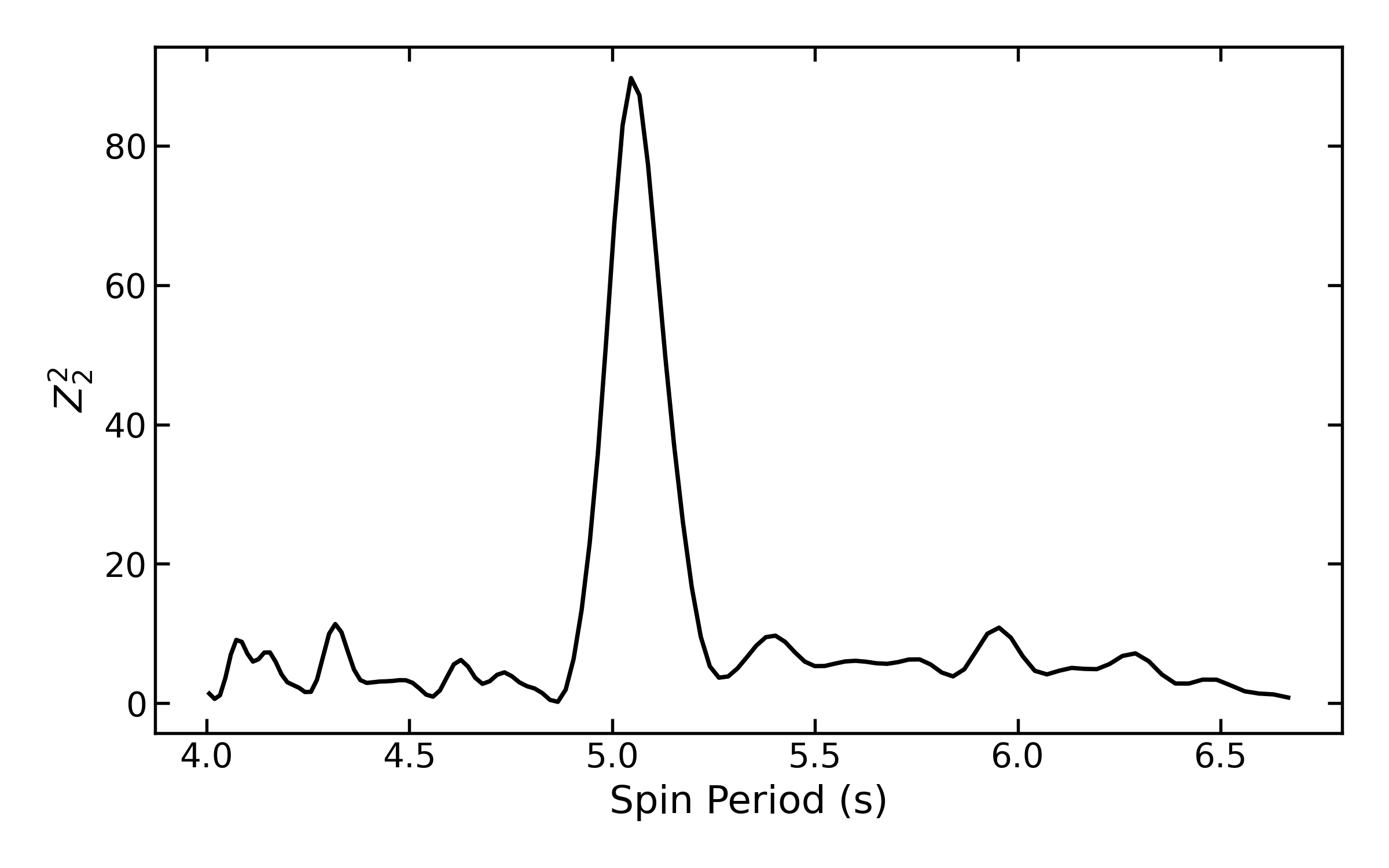} &
\includegraphics[width=0.33\textwidth]{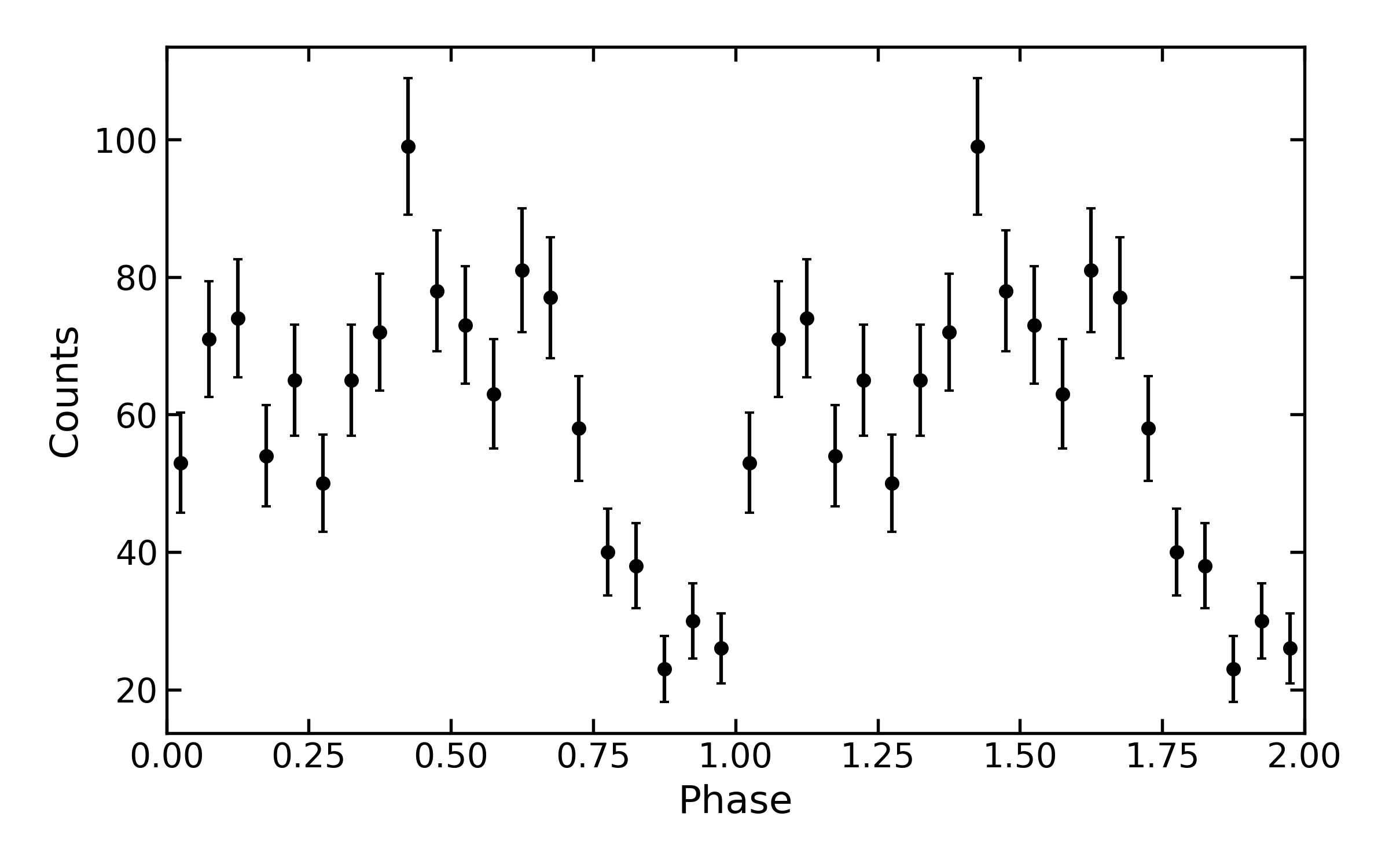} \\
\includegraphics[width=0.33\textwidth]{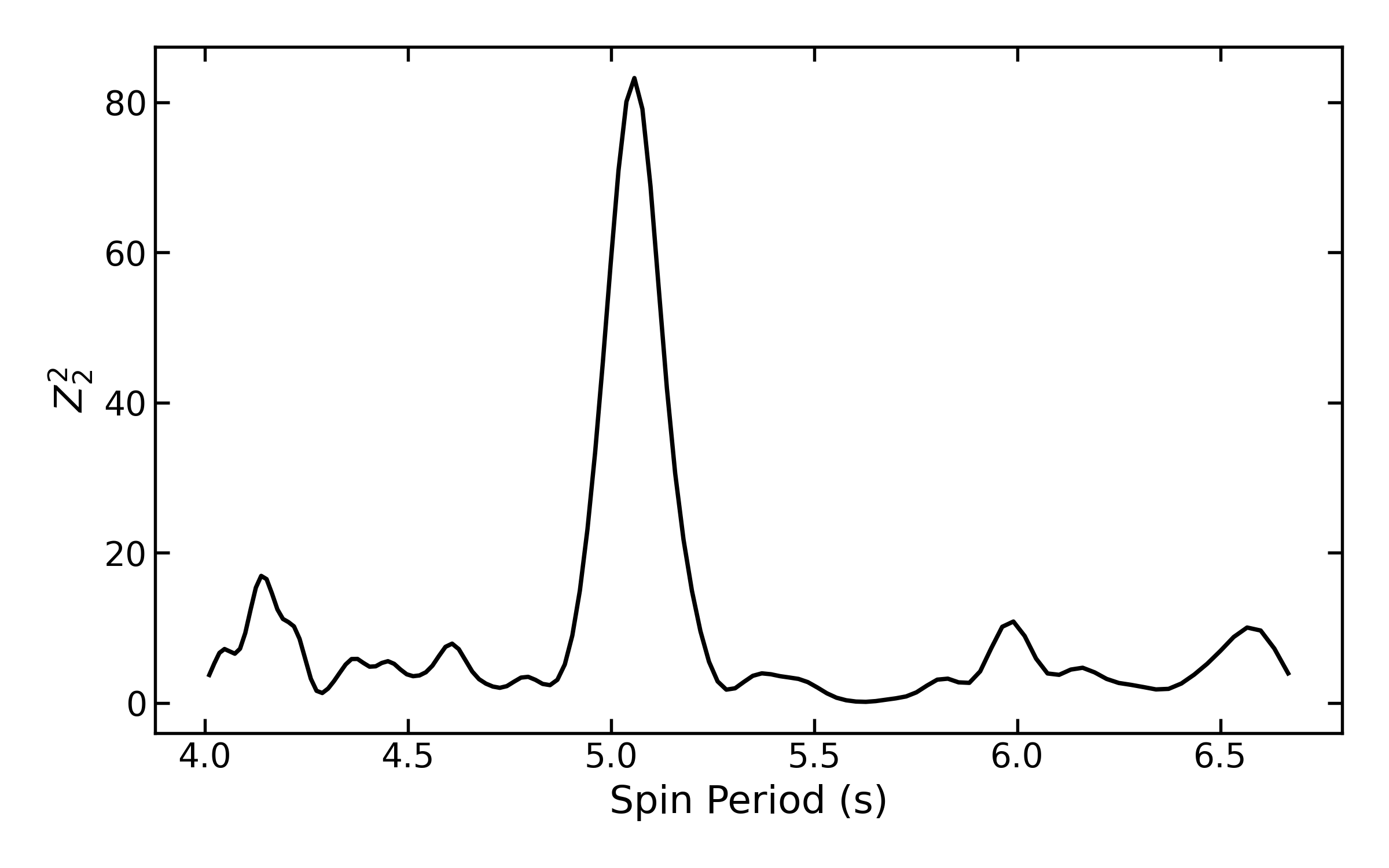} &
\includegraphics[width=0.33\textwidth]{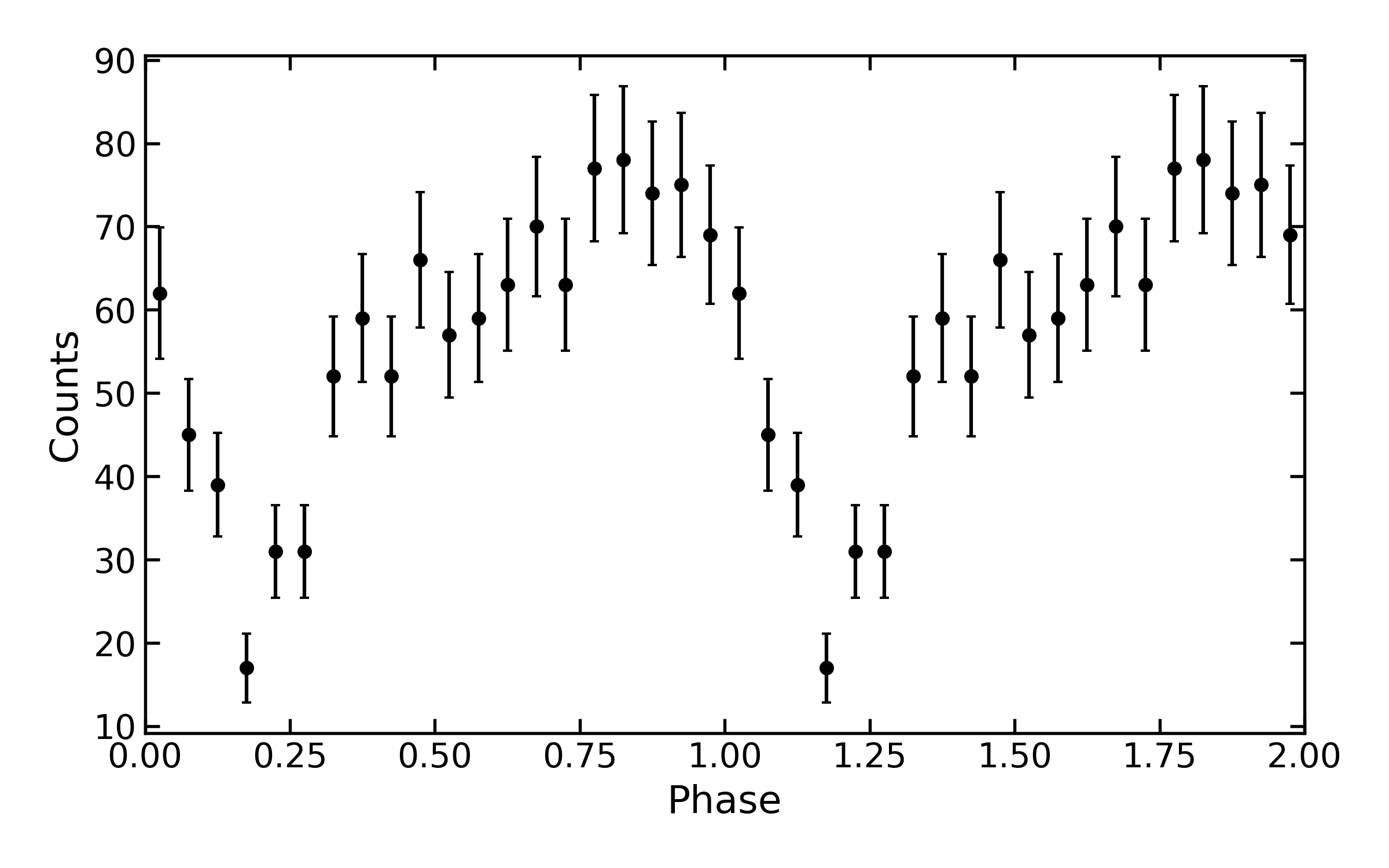} \\
\includegraphics[width=0.33\textwidth]{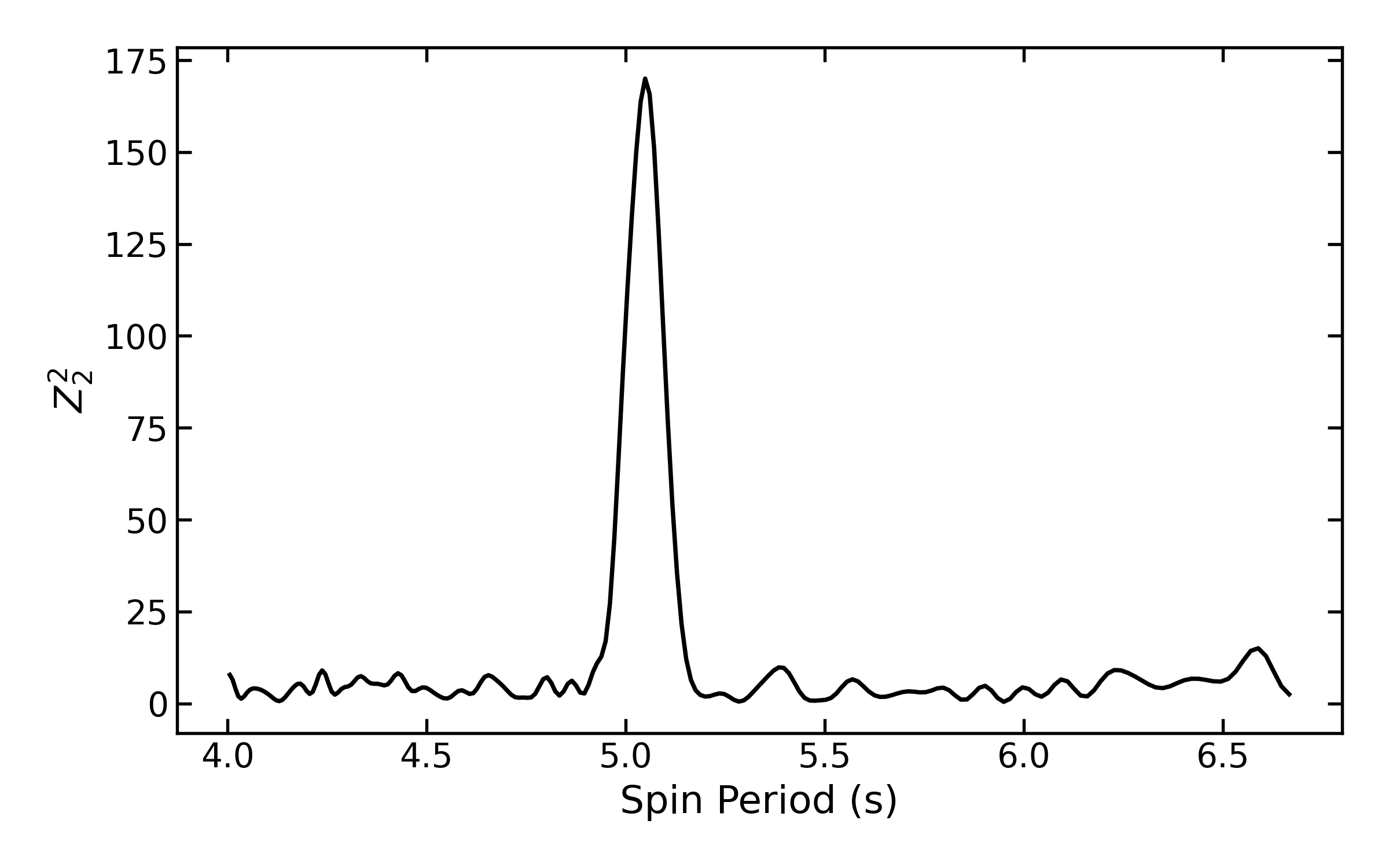} &
\includegraphics[width=0.33\textwidth]{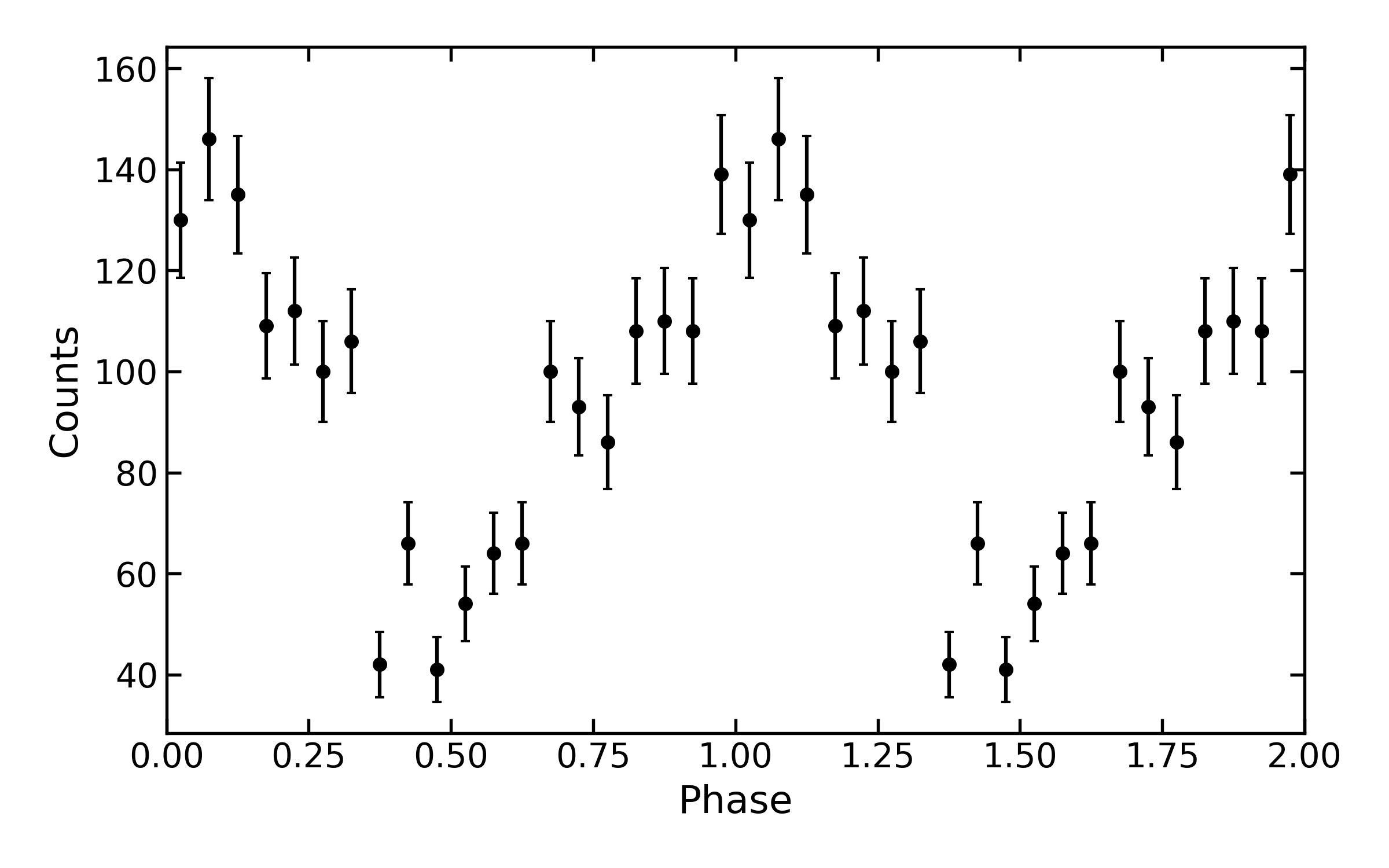} \\
\includegraphics[width=0.33\textwidth]{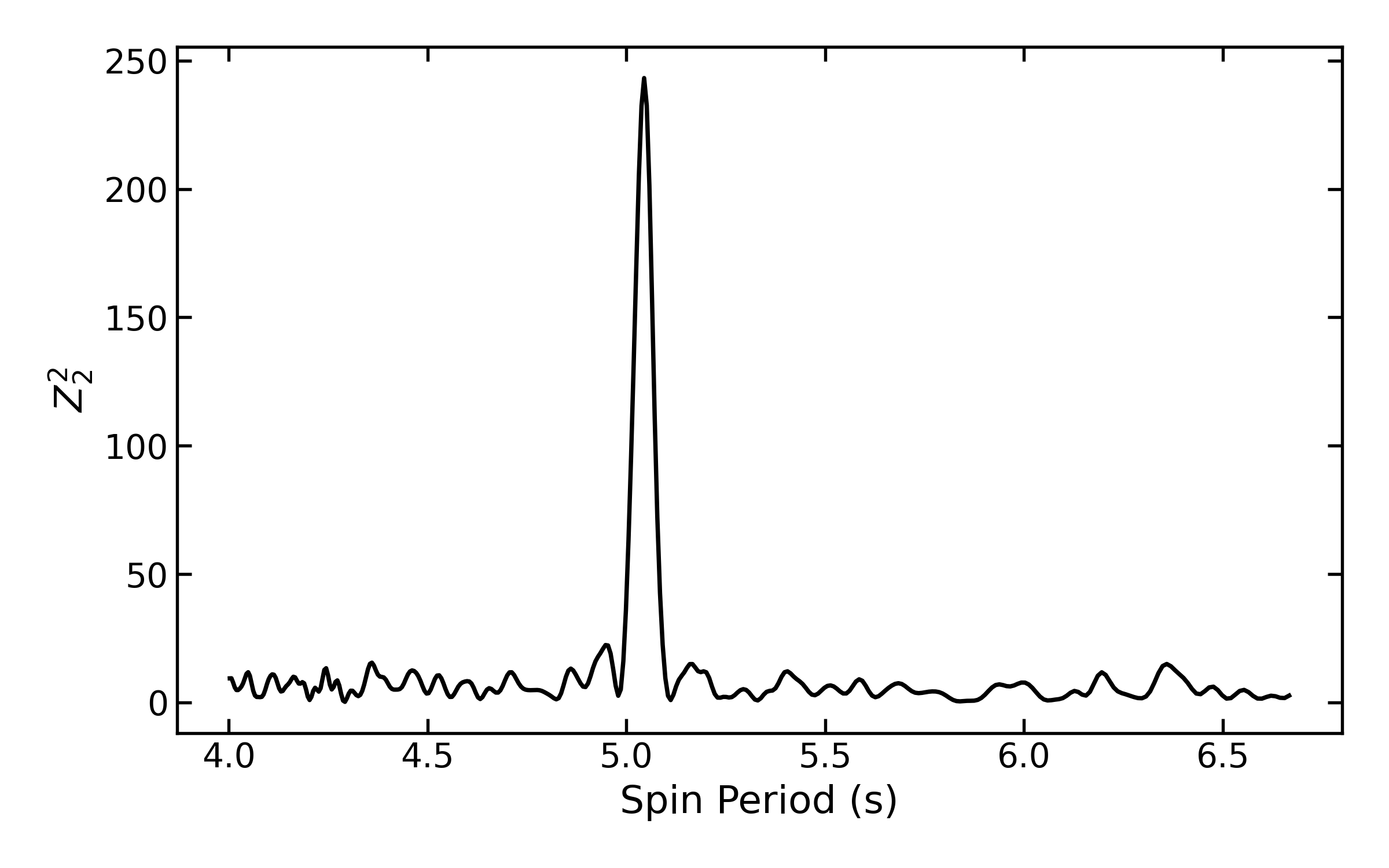} &
\includegraphics[width=0.33\textwidth]{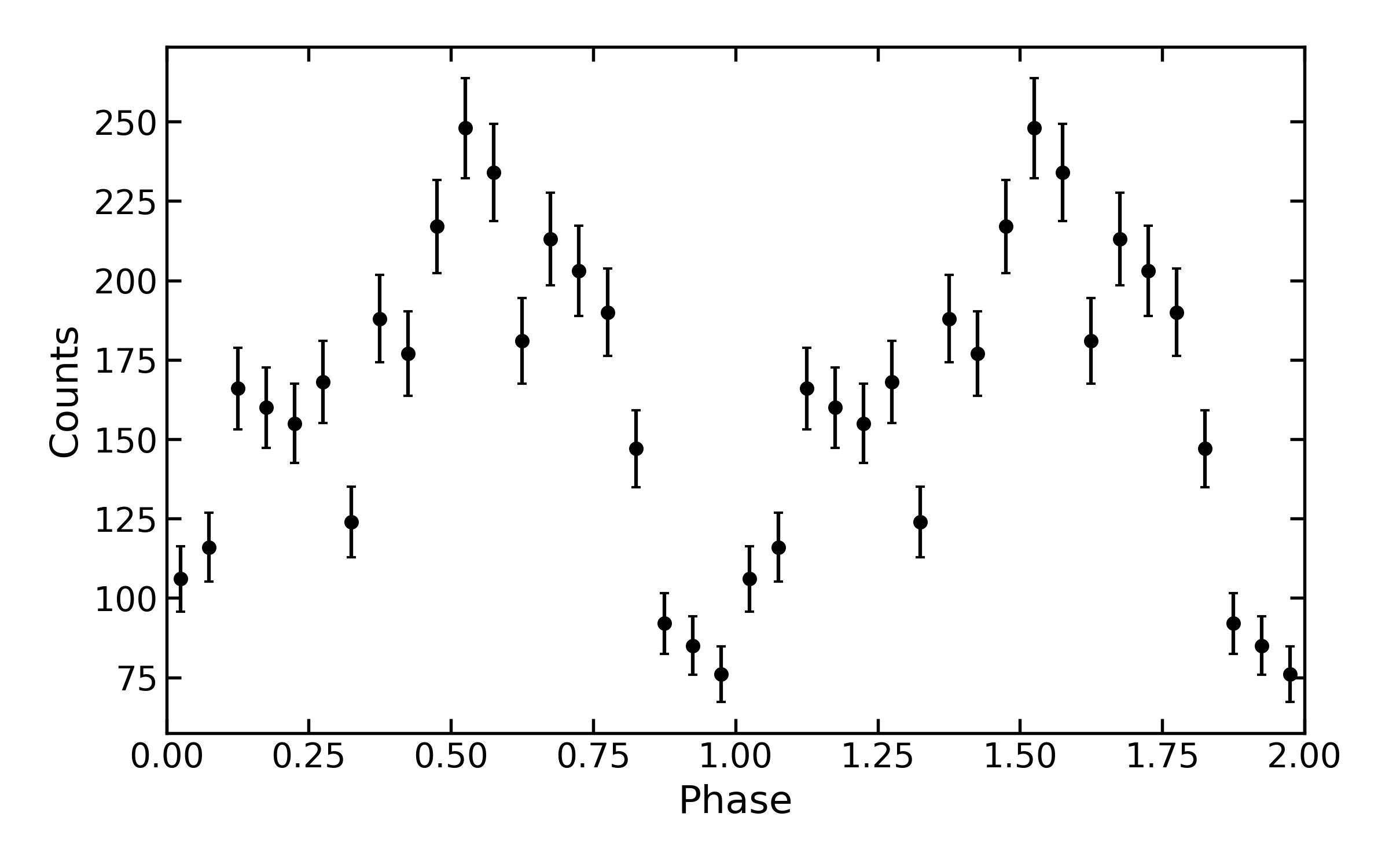} \\
\includegraphics[width=0.33\textwidth]{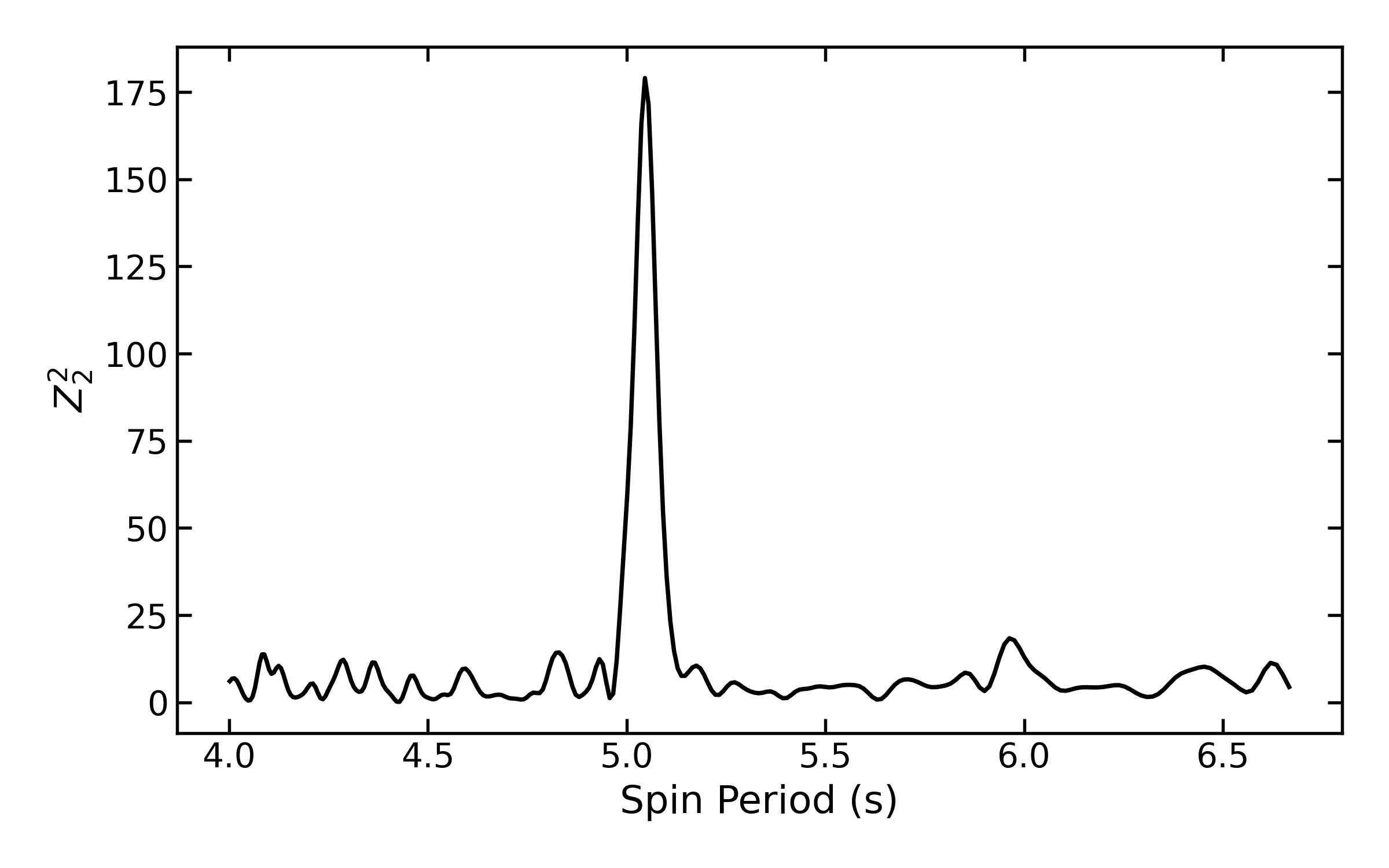} &
\includegraphics[width=0.33\textwidth]{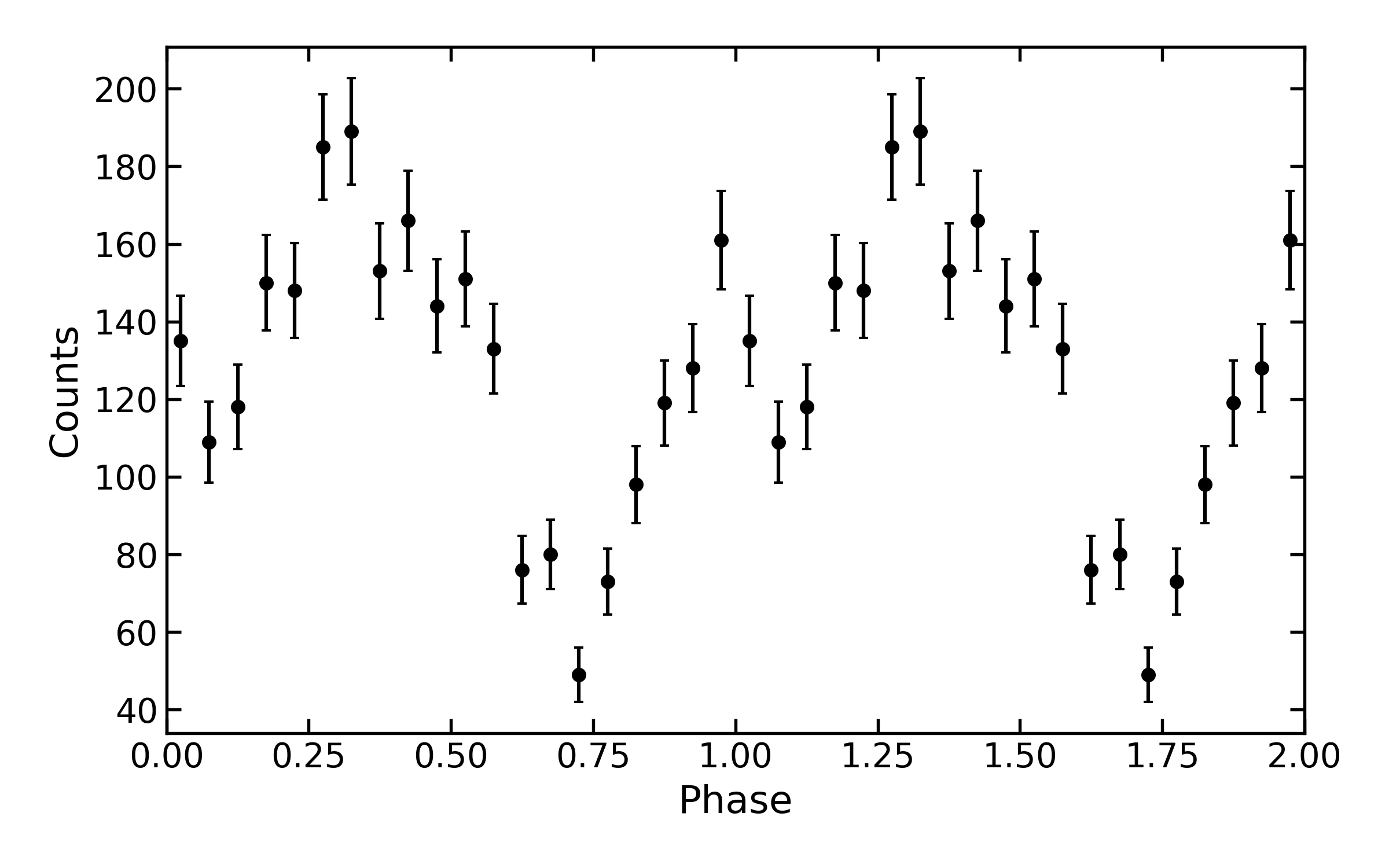} \\
\includegraphics[width=0.33\textwidth]{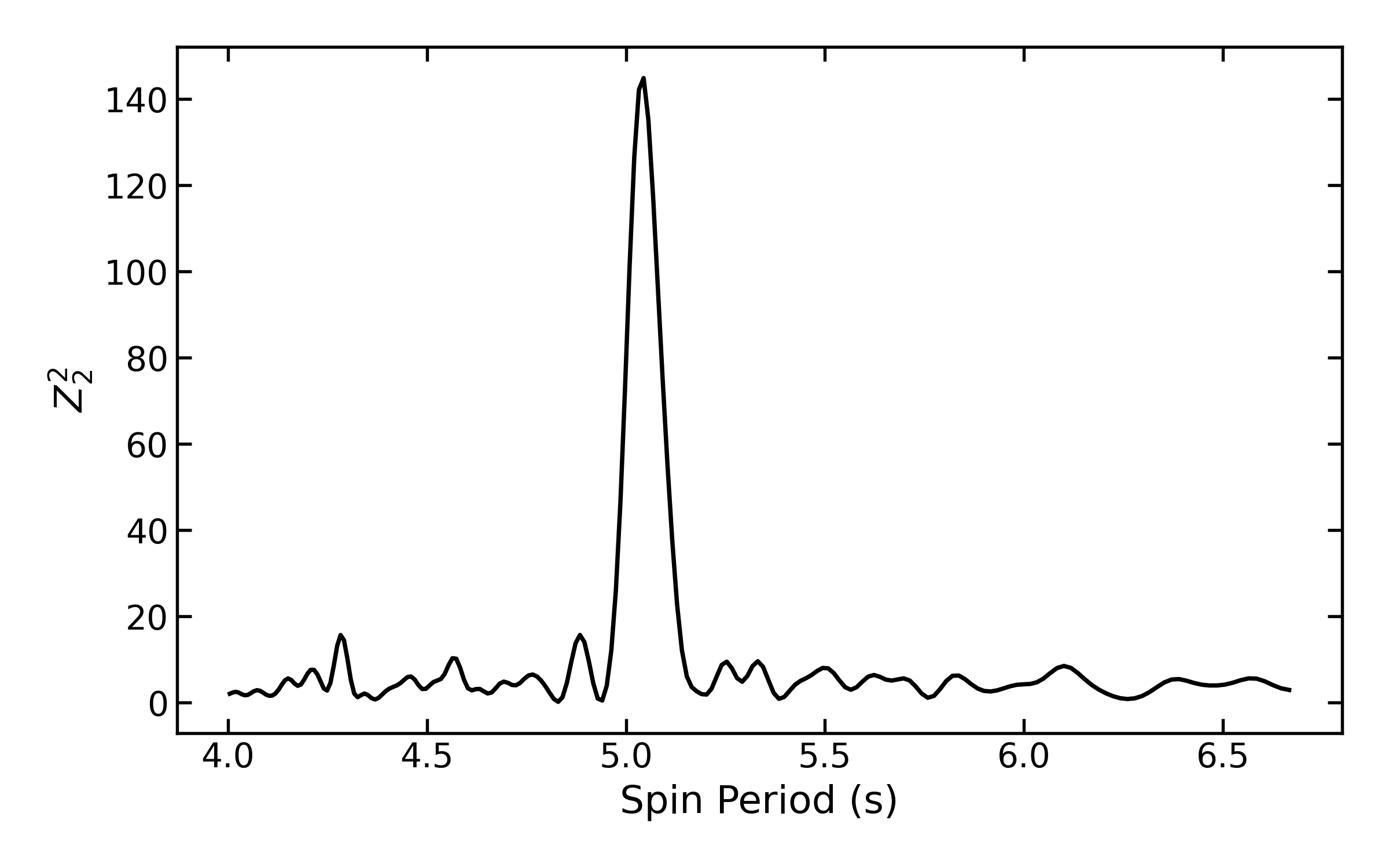} &
\includegraphics[width=0.33\textwidth]{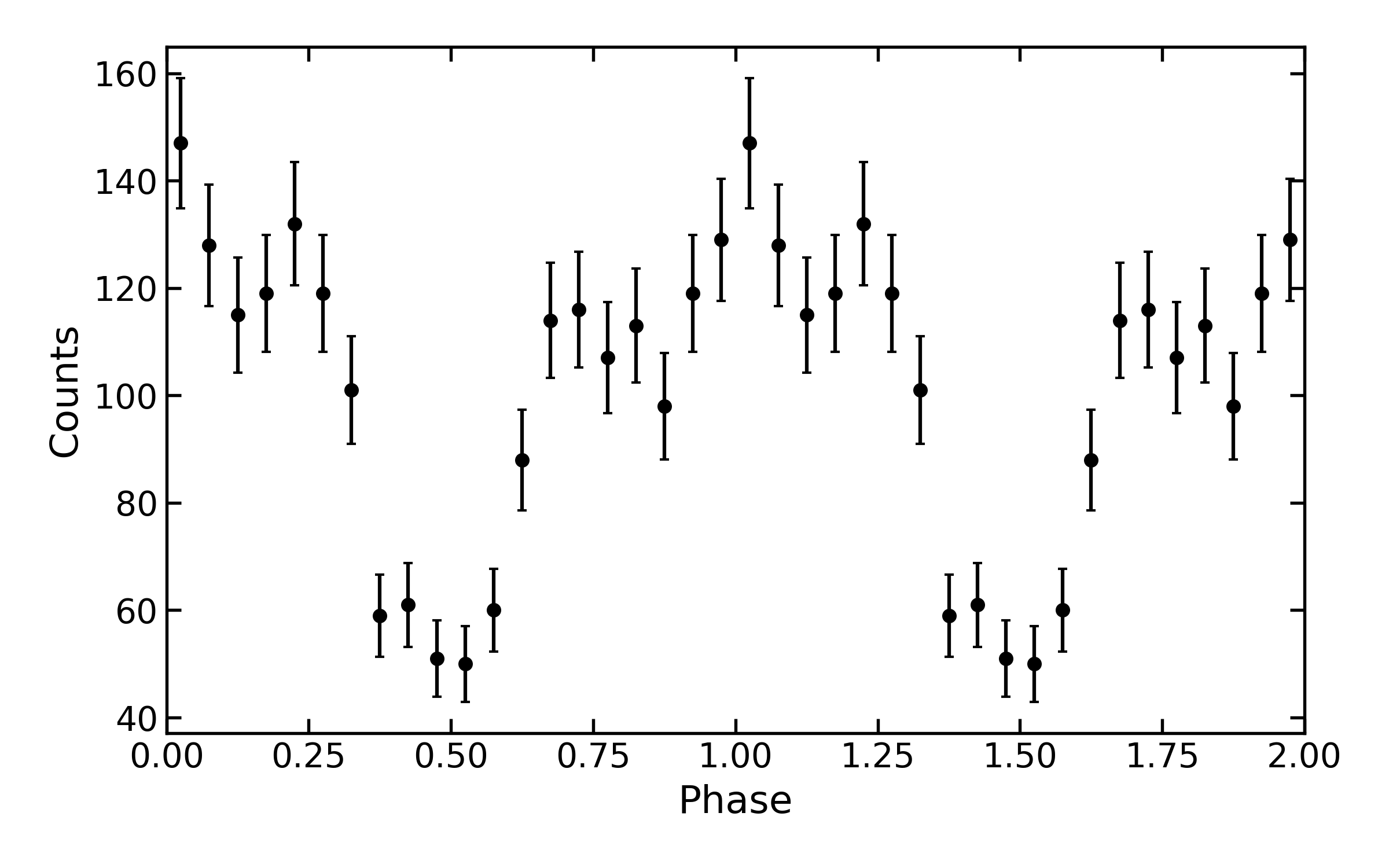} \\
\end{tabular}
\caption{Continued from Fig.~\ref{fig:gti_grid_part1}: power spectra (left) and pulse profiles (right) for ObsID 7204640104, GTI segments 3--8 (rows from top to bottom).}
\label{fig:gti_grid_part2}
\end{figure*}

\begin{figure*}[!htbp]
\centering
\setlength{\tabcolsep}{4pt}
\begin{tabular}{ccc}
\includegraphics[width=0.31\textwidth]{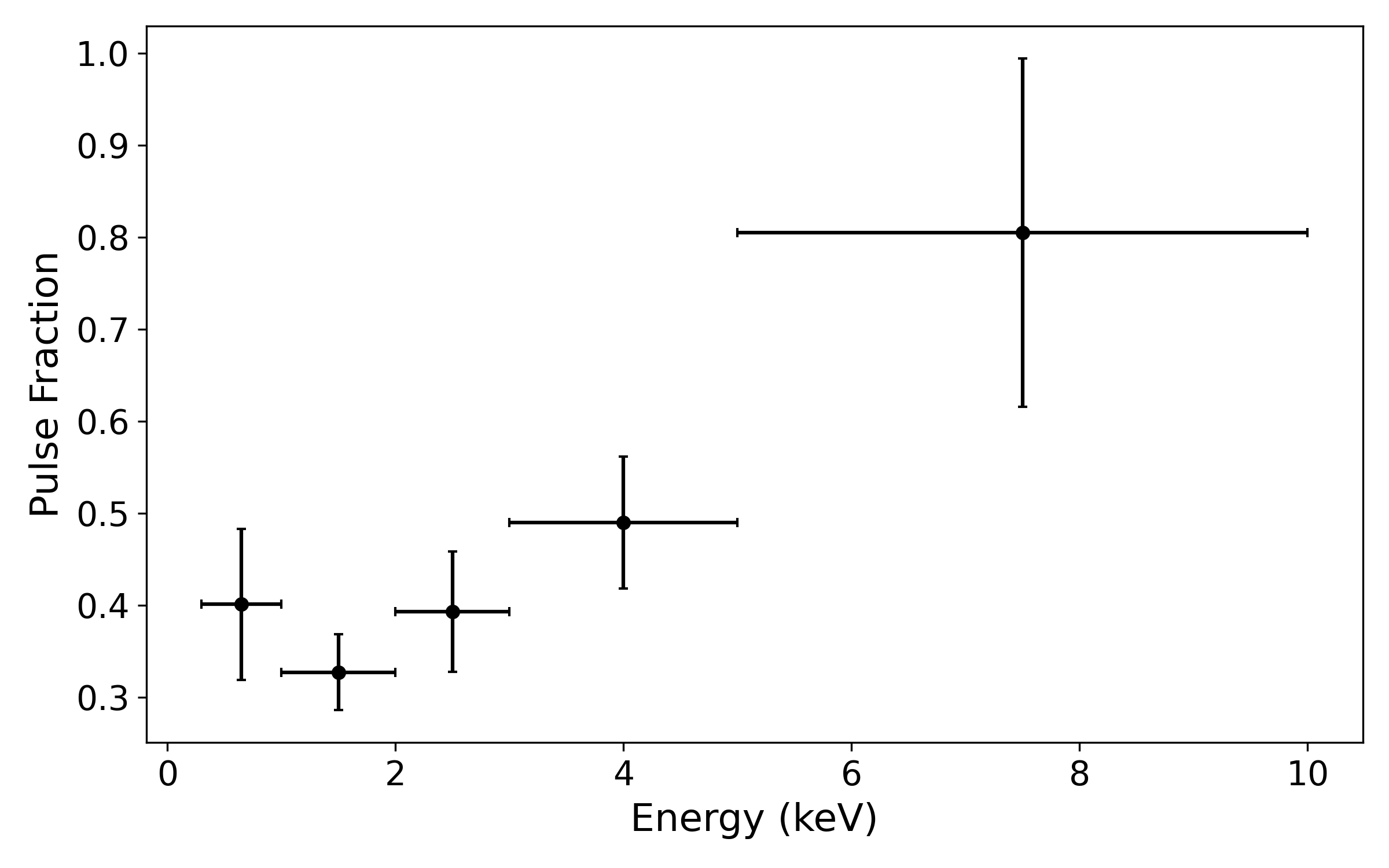} &
\includegraphics[width=0.31\textwidth]{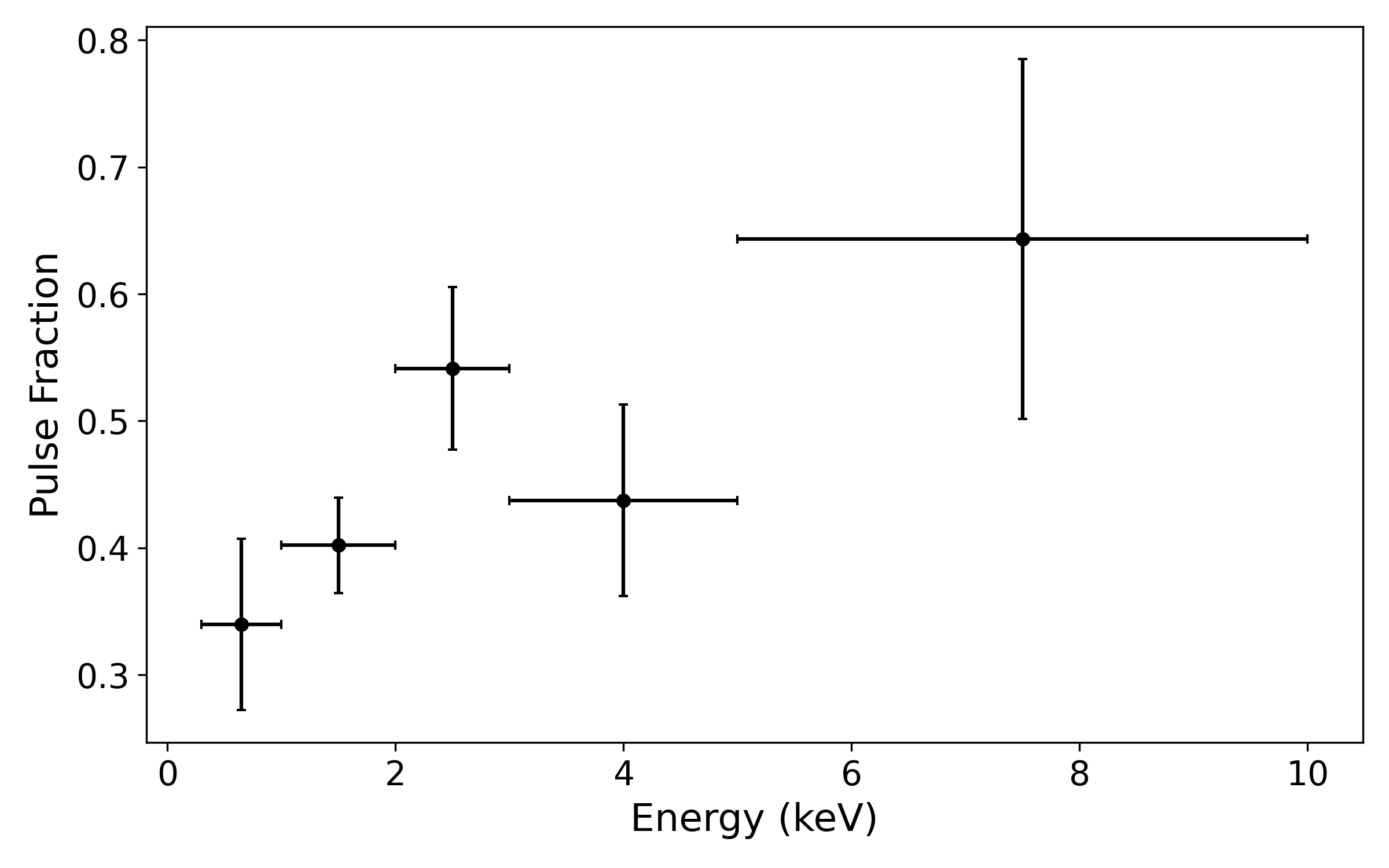} &
\includegraphics[width=0.31\textwidth]{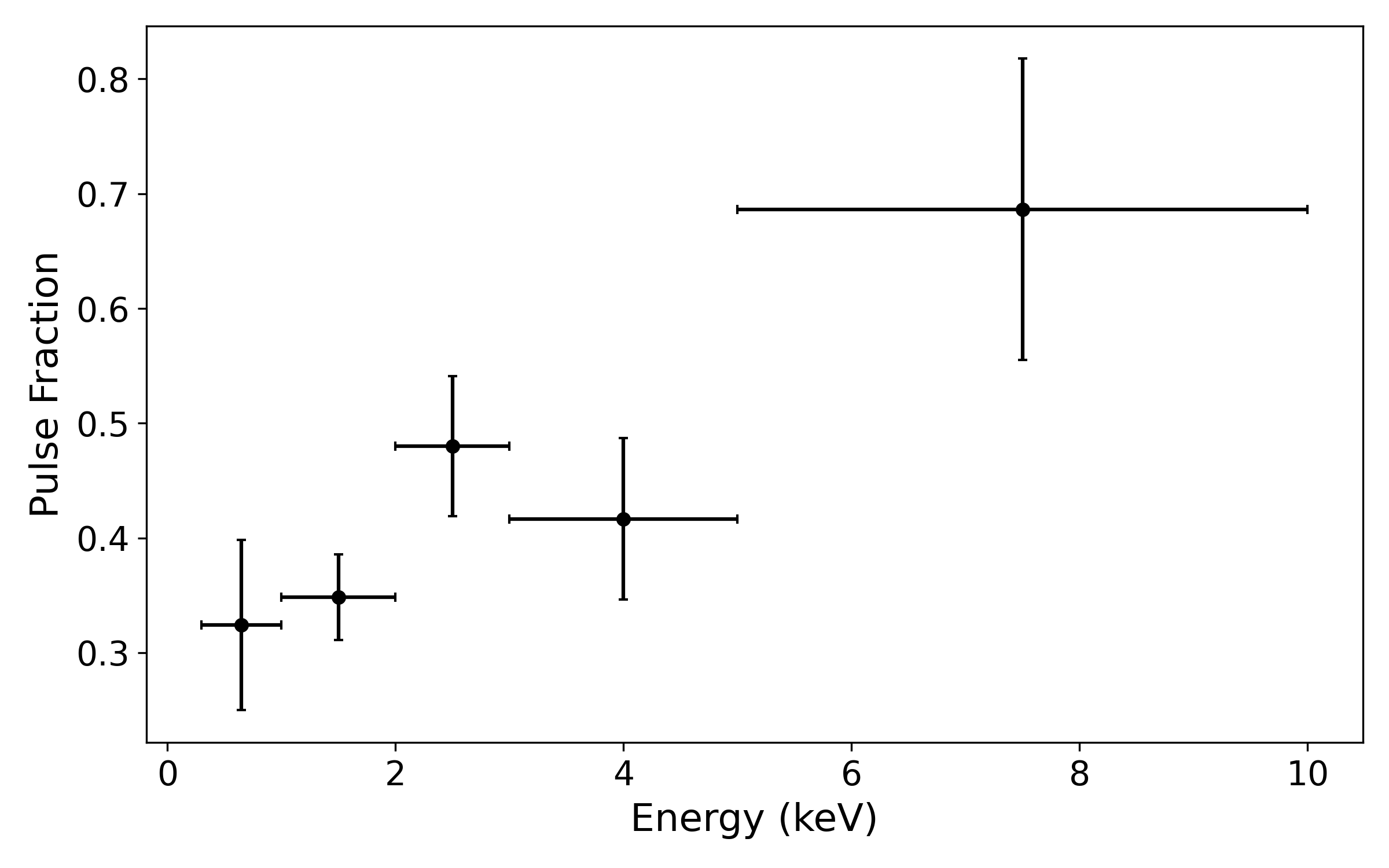} \\
\includegraphics[width=0.31\textwidth]{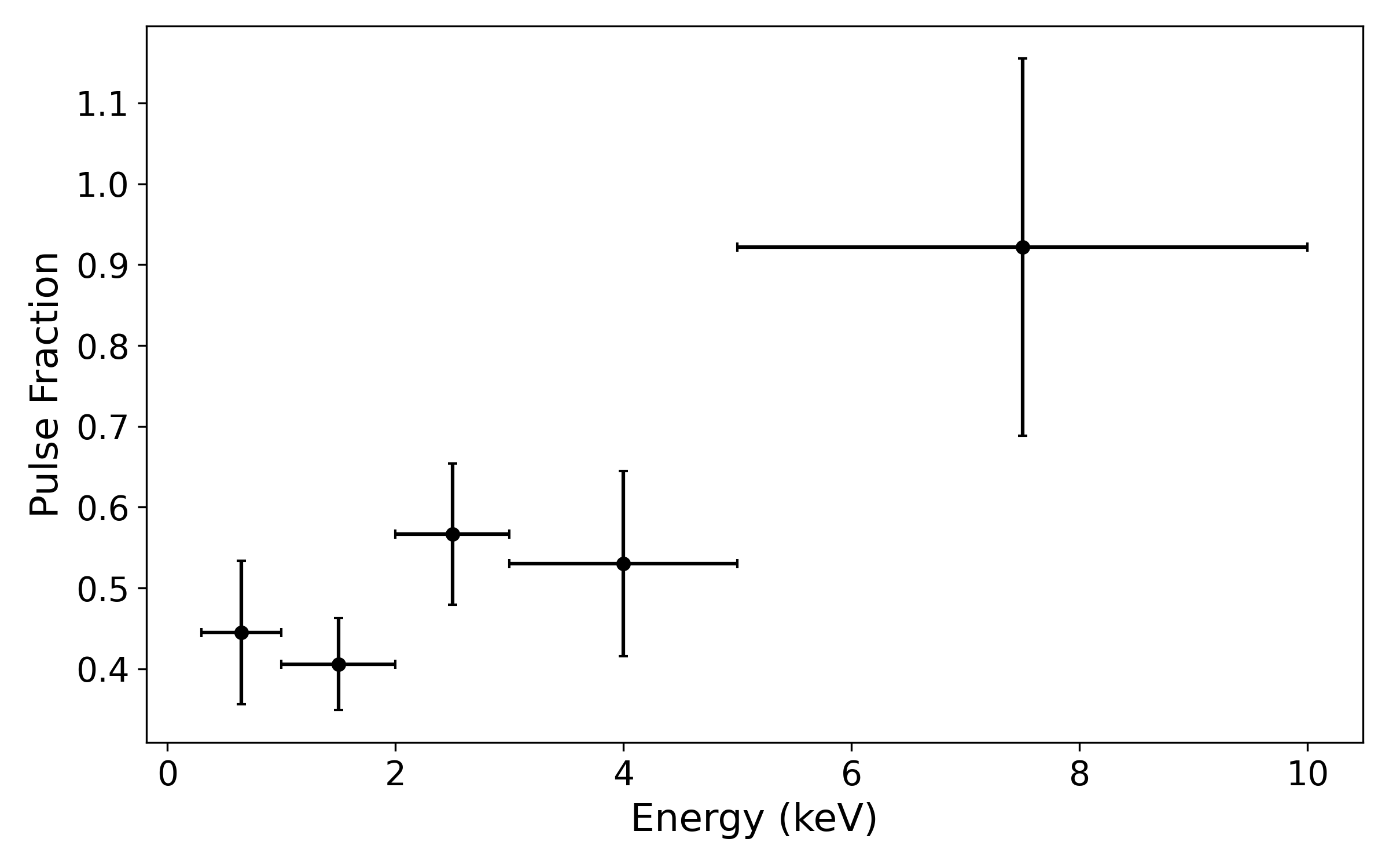} &
\includegraphics[width=0.31\textwidth]{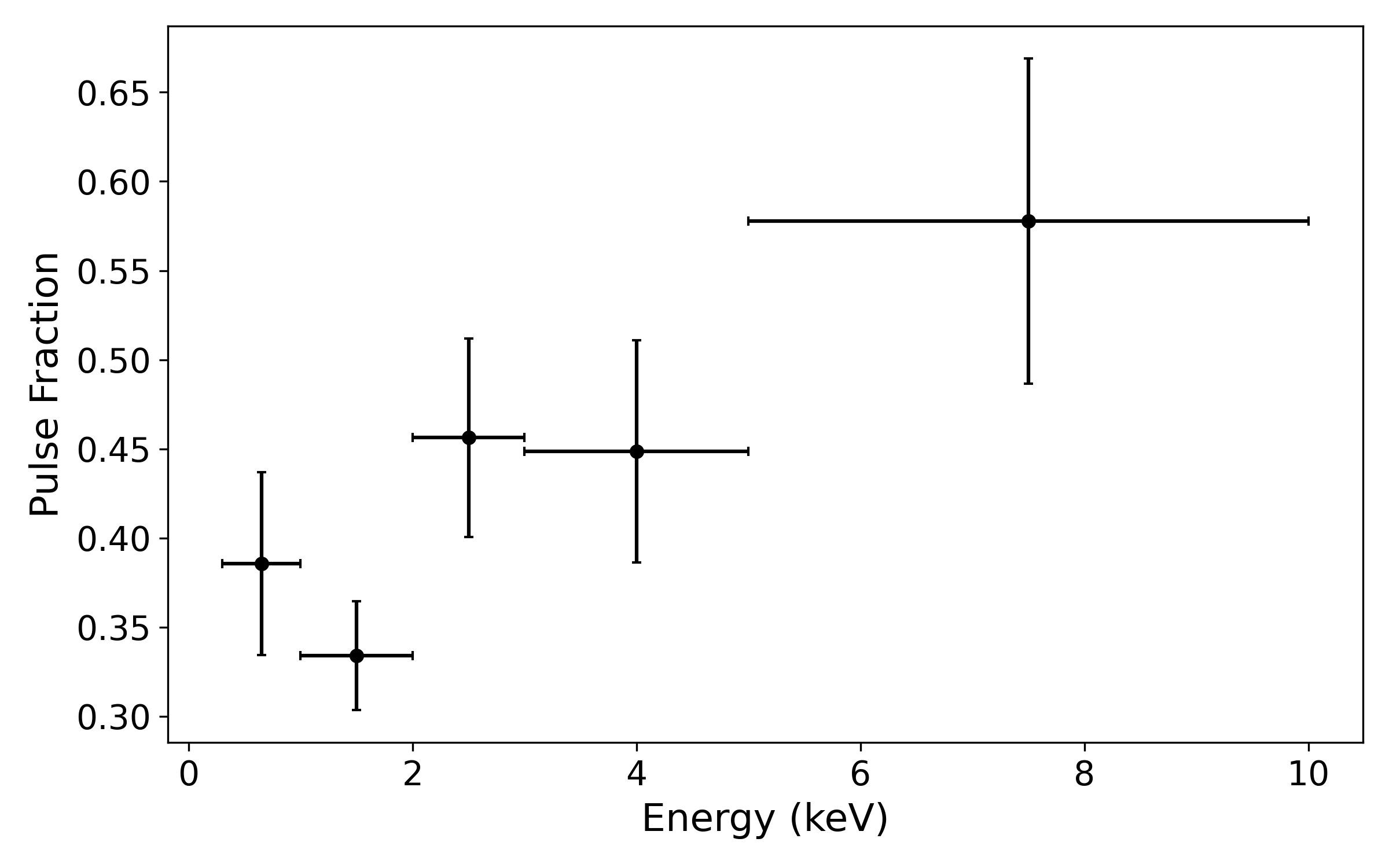} &
\includegraphics[width=0.31\textwidth]{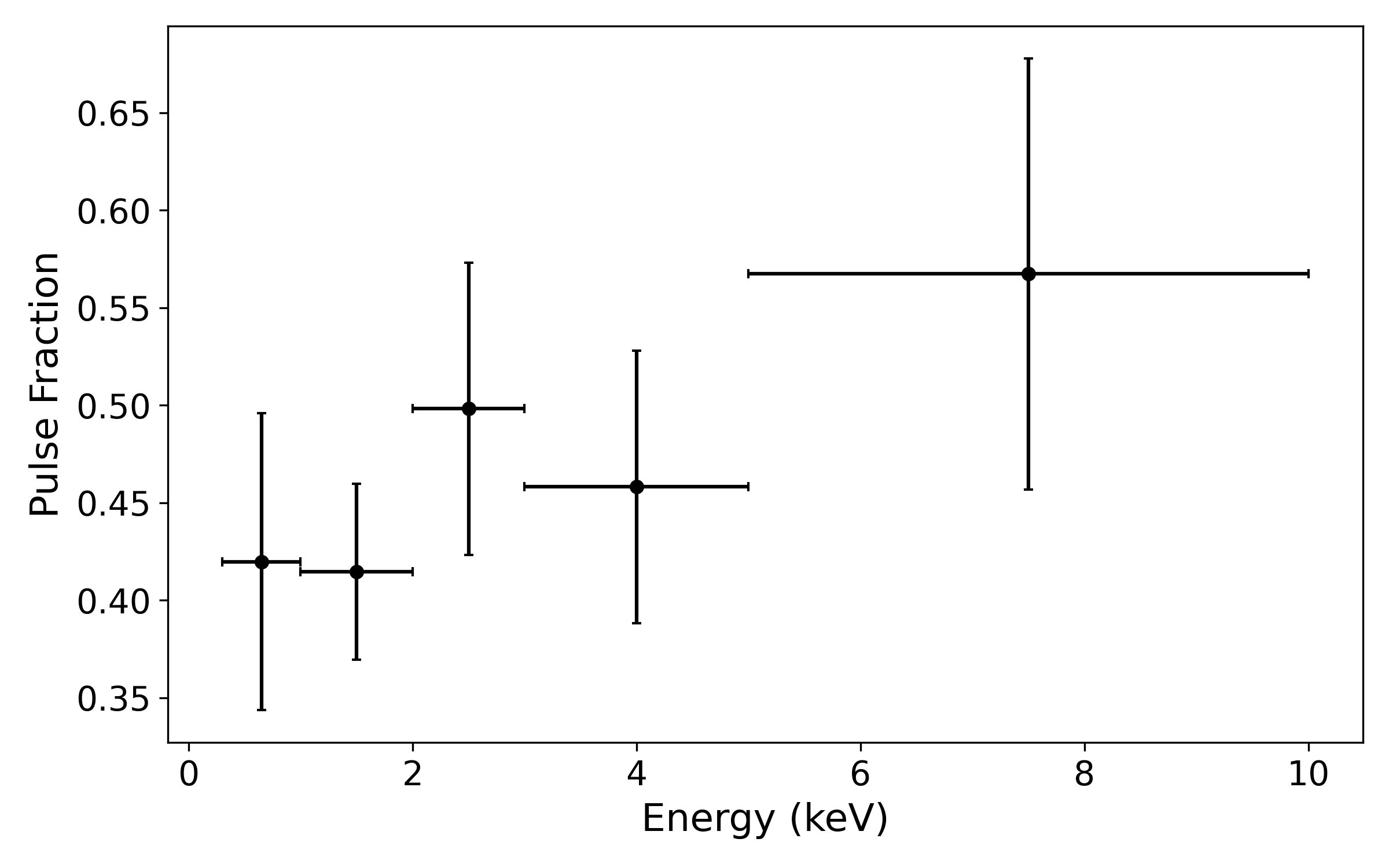} \\
\includegraphics[width=0.31\textwidth]{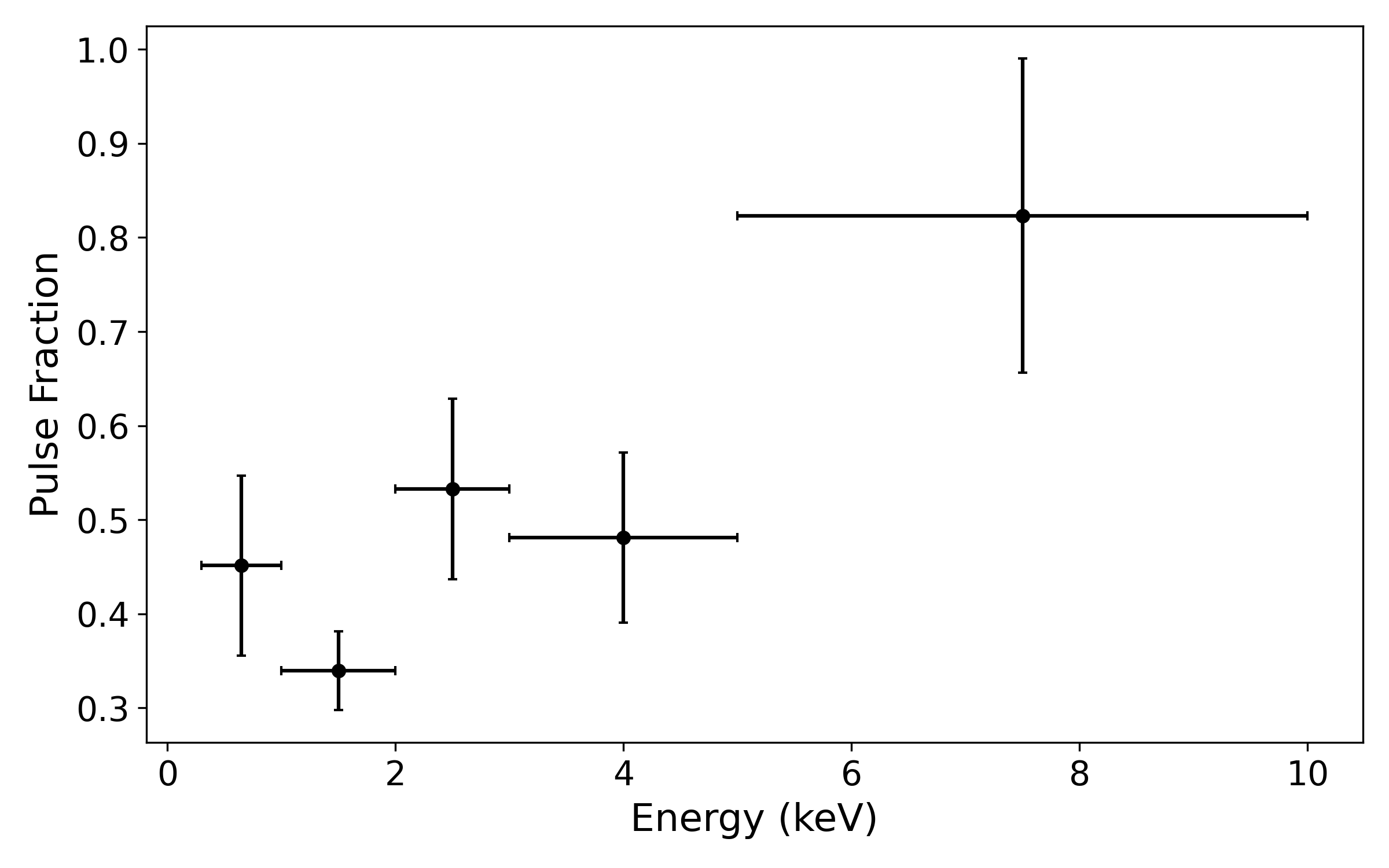} &
\includegraphics[width=0.31\textwidth]{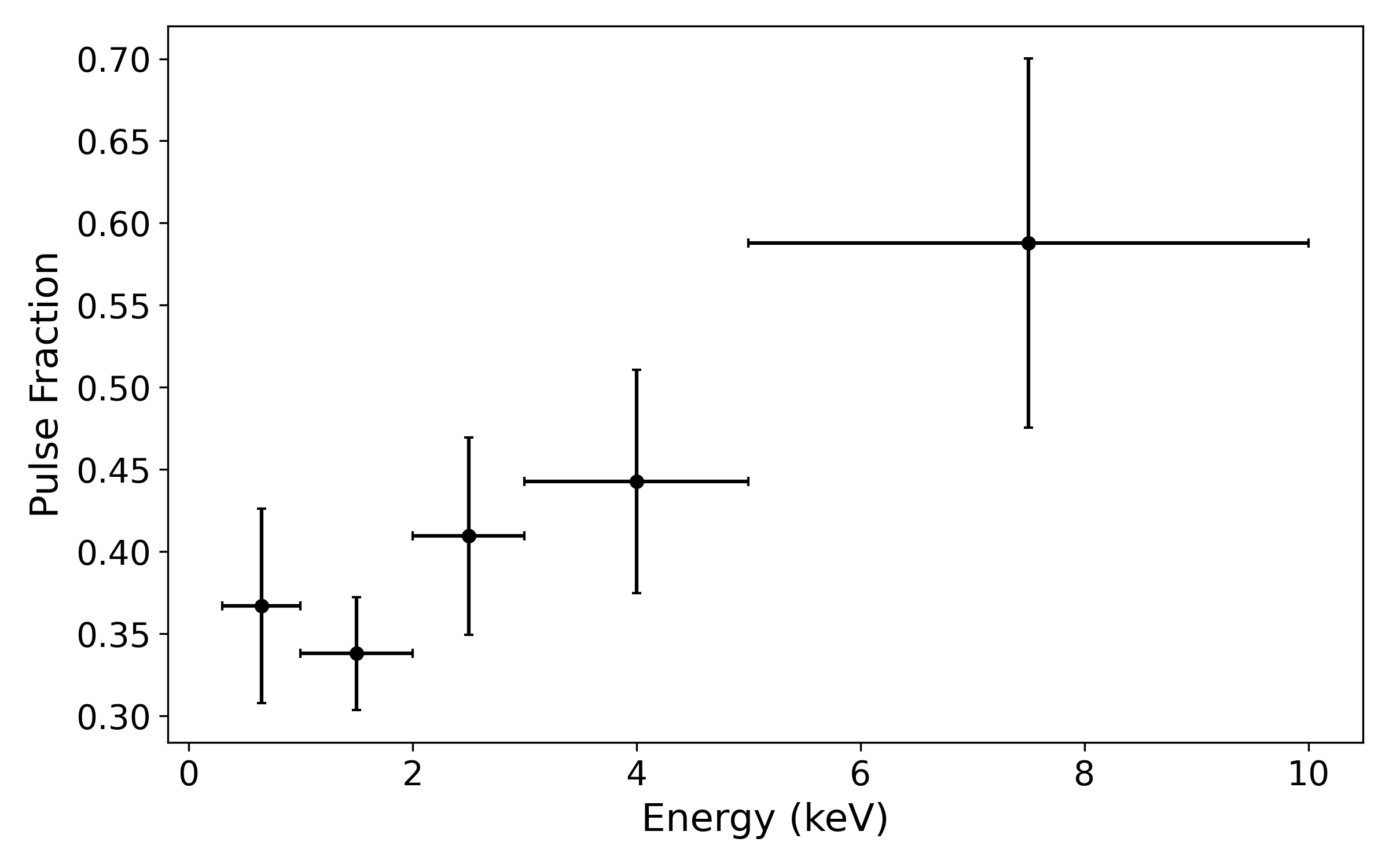} &
\includegraphics[width=0.31\textwidth]{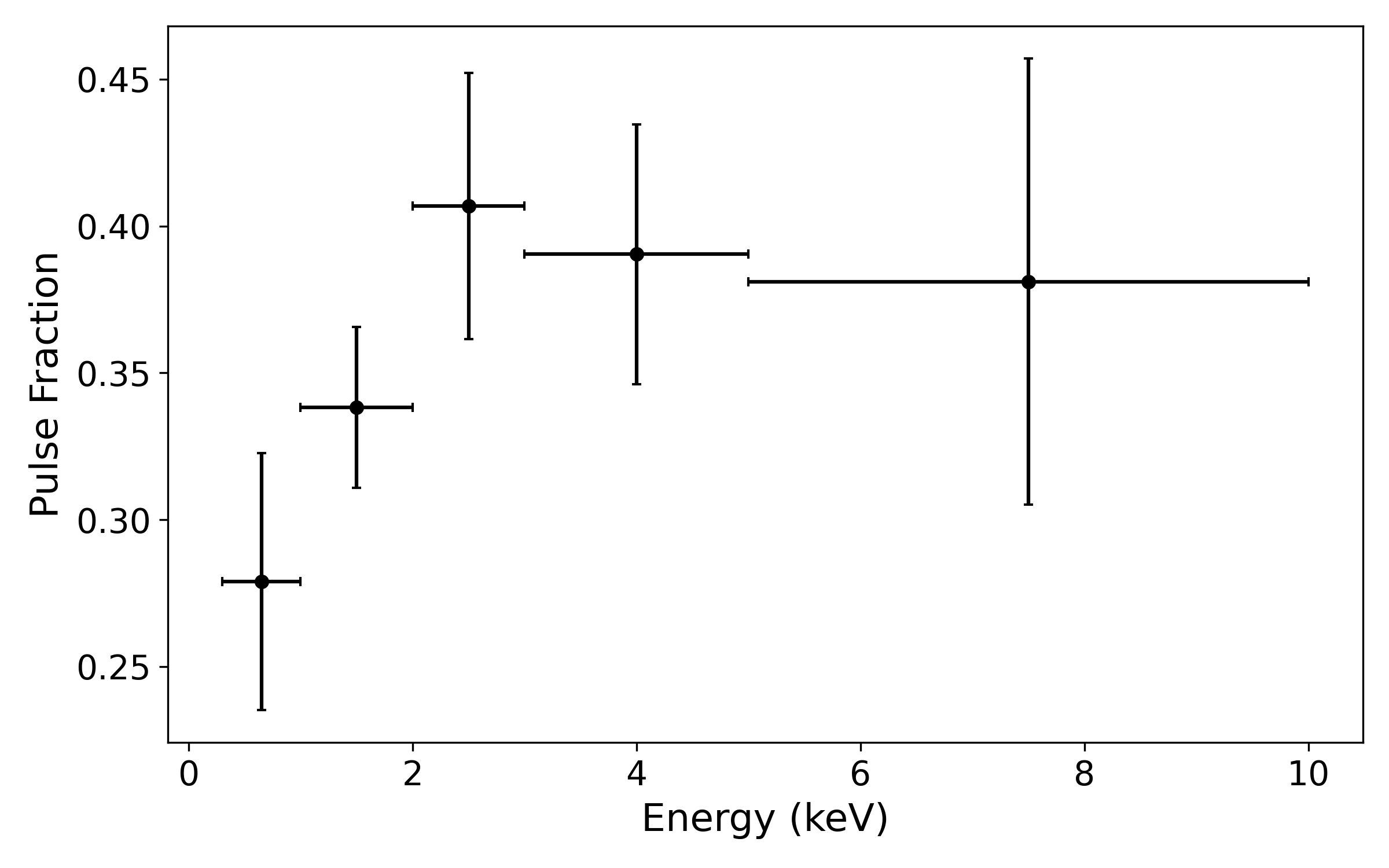} \\
\includegraphics[width=0.31\textwidth]{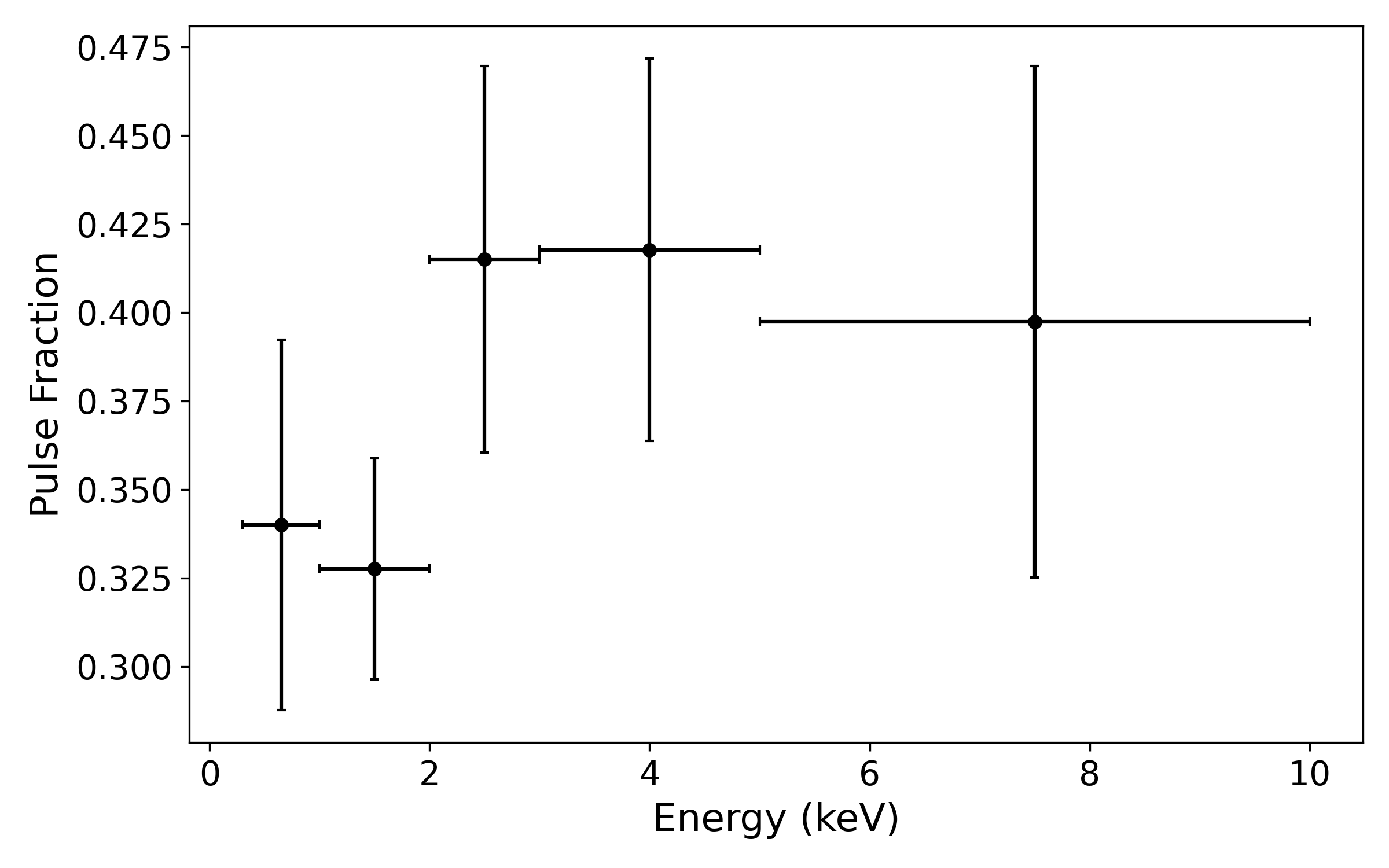} &
\includegraphics[width=0.31\textwidth]{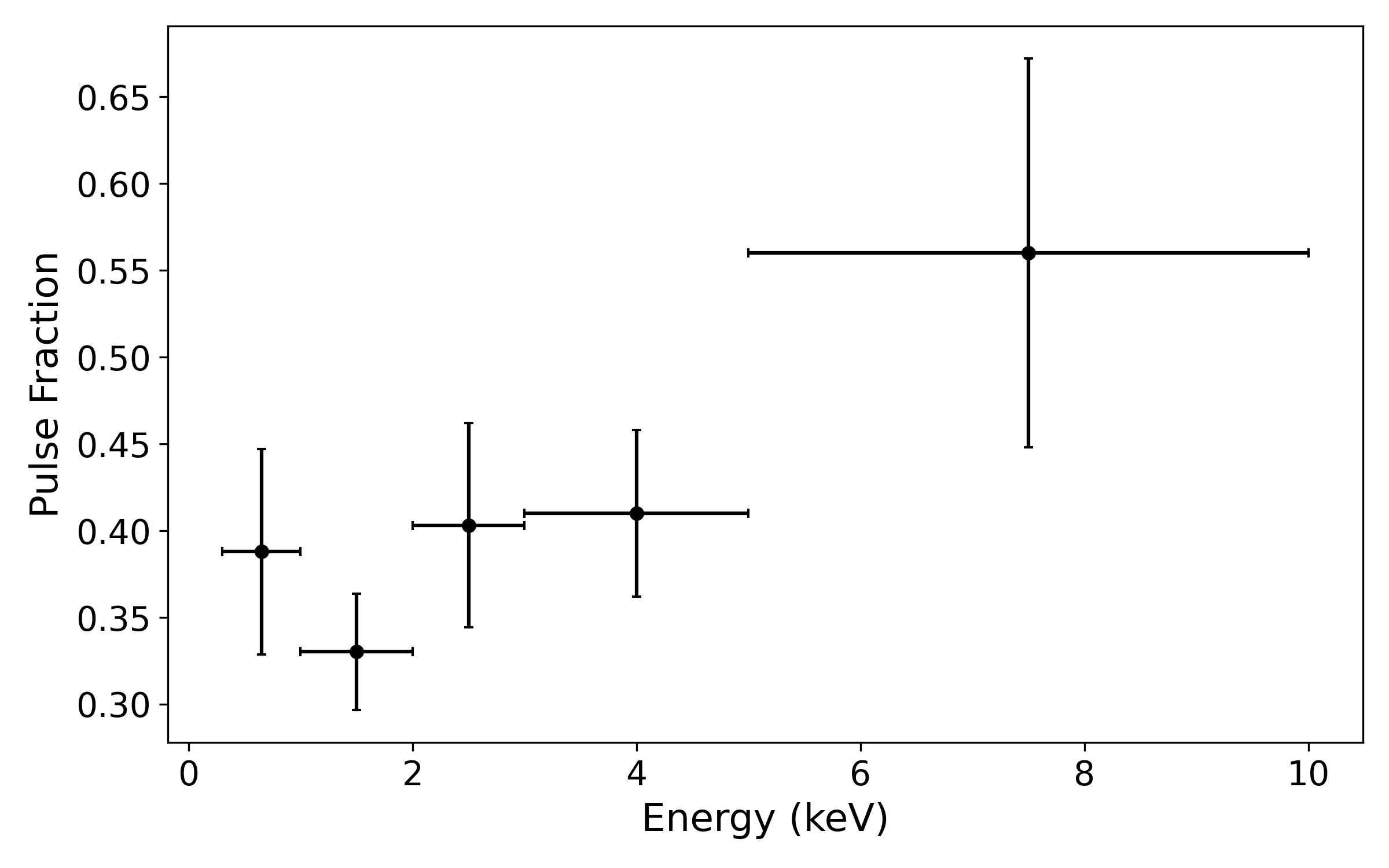} &
\\
\end{tabular}
\caption{Energy dependence of the pulse fraction for individual GTI segments. The first three panels correspond to ObsID 7204640103 (GTI 1--3), followed by ObsID 7204640104 (GTI 1--8), read left to right and top to bottom.}
\label{fig:pf_energy_all}
\end{figure*}

\end{document}